\documentclass[a4paper,11pt]{article}
\pdfoutput=1 

\usepackage{jheppub} 

\usepackage[T1]{fontenc} 
\usepackage{amssymb, bm, graphicx, graphics, color, mathrsfs, multirow}
\usepackage{subfigure} \subfiguretopcaptrue

\newcommand{\be}{\begin{equation}}
\newcommand{\ee}{\end{equation}}
\newcommand{\ben}{\begin{eqnarray}}
\newcommand{\een}{\end{eqnarray}}

\newcommand{\la}{{\lambda}}

\newcommand{\cA}{{\cal A}}

\newcommand{\cL}{{\cal L}}

\newcommand{\p}{\partial}
\newcommand{\na}{\nabla}

\newcommand{\ga}{\gamma}

\makeatletter
\newcommand{\raisemath}[1]{\mathpalette{\raisem@th{#1}}}
\newcommand{\raisem@th}[3]{\raisebox{#1}{$#2#3$}}
\makeatother

\title{\boldmath Static configurations and evolution of higher dimensional brane-dilaton black hole system}

\author[a,b]{Anna Nakonieczna,}
\author[b]{{\L}ukasz Nakonieczny,}
\author[c]{Rafa{\l} Moderski,}
\author[a]{Marek Rogatko}

\affiliation[a]{Institute of~Physics, Maria Curie-Sk{\l}odowska University, \\
Plac Marii Curie-Sk{\l}odowskiej 1, 20-031 Lublin, Poland}
\affiliation[b]{Institute of~Theoretical Physics, Faculty of~Physics, University of~Warsaw, \\
ul. Pasteura 5, 02-093 Warsaw, Poland}
\affiliation[c]{Nicolaus Copernicus Astronomical Center, Polish Academy of Sciences, \\
ul. Bartycka 18, 00-716 Warsaw, Poland}

\emailAdd{aborkow@kft.umcs.lublin.pl}
\emailAdd{Lukasz.Nakonieczny@fuw.edu.pl}
\emailAdd{rogat@kft.umcs.lublin.pl}
\emailAdd{moderski@camk.edu.pl}

\abstract{
Static configurations and a dynamical evolution  of the system composed of a higher-dimensional spherically symmetric dilaton black hole and the Dirac-Goto-Nambu brane were investigated.
The studies were conducted for three values of the dilaton coupling constant, describing the uncoupled case, the low-energy limit of the string theory and dimensionally reduced
Klein-Kaluza theories. 
When the black hole is nonextremal, two types of static configurations are observed, a brane which intersects the black hole horizon and a brane not having any common points with the accompanying black hole.  As the 
number of spacetime dimensions increases, the brane bend in the vicinity of the black hole disappears closer to its horizon. Dynamical evolution of the system results in an expulsion of the black hole from the brane. It proceeds faster for bigger values of the bulk spacetime dimension and thicker branes. The value of the dilatonic coupling constant does not influence neither the static configurations nor the dynamical behavior of the examined nonextremal system.
In the extremal dilaton black hole case one obtains expulsion of the brane which is independent on the spacetime dimensionality and the value of the coupling constant.
Dynamical studies of the configurations in the extremal case reveal that the course of evolution of the system is similar to the nonextremal one, except for a slightly earlier expulsion of the black hole from the brane.
}

\begin{document} 
\maketitle
\flushbottom

\section{Introduction}
\label{sec:intro}
A resurgence of studies of higher-dimensional black holes, black branes and their mutual interactions has been observed recently. These subjects are valid in several different
areas of modern theoretical physics. 
Contemporary candidates of quantum gravity such as the string/M-theory predict that our universe can be described as a brane embedded in a higher-dimensional spacetime.
This idea was incorporated in the so-called brane-world models which are composed of a higher-dimensional bulk and a lower-dimensional embedded brane.
The first brane-world models were built with a tensionless brane and compactified flat extra dimensions~\cite{ark98,ark99}.  Later on, they were extended to the case of a non-zero tension brane and a warped
(negatively curved) spacetime~\cite{ran1,ran2}. The Standard Model particles are confined to the brane, while
gravity can penetrate the other additional dimensions. In such theories interactions between the brane and black holes were intensively investigated~\cite{tan11}.
Especially, much attention was paid to the problem of higher-dimensional black holes/black objects crossing the brane and the expulsion problem for such systems. In~\cite{fro06} a toy model for studying a process 
of merging within the brane-black hole system and a topology change was considered. The higher-dimensional generalizations both in classical general relativity and in certain strongly coupled gauge theories were presented in~\cite{kol05,kol06,mat06,mat07}. Axisymmetric higher-dimensional systems consisting of stationary test branes interacting with Myers-Perry bulk black holes were examined for arbitrary brane and bulk dimensions~\cite{czi13}.

One of the predictions of the theories with large extra dimensions is the conjecture that mini black holes with masses of $1$~TeV can be in principle observed in LHC~\cite{lhc,lhc1}.
Namely, when two particles on the brane collide at the center of mass energy larger than $1$~TeV, the system collapses and the formation of mini black hole begins. 
With the passage of time the black hole 
starts to emit Hawking radiation into lower dimensional fields localized on the brane and partly into the higher-dimensional mode, i.e., into the bulk.

The early stages of the evolution of our Universe are also the arenas where primordial black holes and domain walls could have been born. Cosmological density
perturbations may have collapsed and led to the formation of small black holes. Next, when the Universe underwent several cooling
processes series of phase transitions might occurred leading to the domain wall and other cosmological defects formation~\cite{vil}. In principle, the interaction of 
black holes with branes (domain walls) should be taken into account when discussing the aforementioned period of the Universe evolution.

In the past few decades, the interactions between black holes and domain walls have been an object of serious studies, both analytical and numerical.
A Nambu-Goto membrane in the Reissner-Nordstr\"om-de Sitter spacetime was studied in~\cite{hig00}, while the gravitational system of the Schwarzschild static black hole and a thick domain
wall was examined in~\cite{mor00,mor03}. The domain walls were simulated by a scalar field with an adequate potential like $\phi^4$ or the sine-Gordon ones. 
The dilaton black hole-domain wall system was elaborated in~\cite{rog01,mod03}. Among all, it was shown that in the extremal case the expulsion of the scalar field connected with a domain wall
took place. In~\cite{mod04} it was revealed that there was a parameter depending on the black hole mass and the width of a domain wall, which turned out to be an upper limit for the expulsion process.
The thick domain wall problem in the anti-de Sitter spacetime was treated in~\cite{mod06}, while the cosmological brane-black hole system in~\cite{rog04}. By using the C-metric construction the line element describing
an infinitely thin brane intersected by a cosmological black hole was derived.

In~\cite{fla05} the black hole whose size was small when compared to the extra dimensions was considered (the tension of the brane was negligible). The simulation was conducted 
by looking for the dynamical evolution of the brane in the fixed background of the black hole. Further studies of the interaction of a black hole with a scalar field domain wall confirm the previously drawn conclusions~\cite{fla06a}. 
The evolution of the system was also analyzed during the separation process.
In~\cite{fla06b} the problem of the influence of the brane tension on the escaping black hole was elaborated. The new challenge was to find if the brane tension might 
prevent the black hole from escaping for a small recoil velocity. The numerical studies of the interaction of black holes with the field theoretical domain wall, i.e., 
described by an axion-like field theory with an approximate $U(1)$-gauge symmetry, exploring the topology changing process of a brane perforation was conducted in~\cite{fla07}.
It was revealed that the perforation process depends on the collision velocity and there was a critical value of velocity which suppressed the domain wall perforation.

In~\cite{cha00,sto05} the studies of the domain walls perforation by black holes envisaged that the mechanism is more or less relevant to the cosmological domain wall case.
On the other hand, the curvature corrections to the static axisymmetric membrane evolution in the spherically symmetric spacetime of an arbitrary dimensional black hole, were given
in~\cite{czi09}. The problem of a thick brane-black hole system was explored in a series of papers, where perturbative and non-perturbative solutions including the case of a two-brane were studied 
\cite{czi09,czi10,czi11}. The last moments of a mini black hole escaping from a brane were examined in~\cite{bal11}. Among all, it was shown that the world sheet becomes isotropic at the reconnection point. It turns out that the
higher dimension we examine, the faster the brane becomes isotropic.

A collision of a point particle with an infinitely thin planar domain wall interacting within the linearized gravity in the Minkowski spacetime of an arbitrary dimension was analyzed in~\cite{gal14}. The energy momentum balance in this system was found in~\cite{gal14a}.

The motivation standing behind our studies is the fact that the classical black hole solutions
of Einstein-Maxwell gravity in four or higher dimensions have quite new features
when the underlying theory is modified by the introduction of the low-energy string corrections.
The key property of the aforementioned correction in the action is connected with the nontrivial 
coupling of the dilaton field (as well as the other ones like, e.g., axions) with the field strength
of the gauge field or gauge fields. There are some models which admit $n$-gauge fields.

In our paper we shall consider static configurations and a dynamical evolution of the higher dimensional
spherically symmetric black hole and the Dirac-Goto-Nambu brane (a brane-dilaton black hole (BDBH) system),
paying attention to the new features of the system in question. 
The dilaton black hole was briefly described in section \ref{sec:bulk} to 
establish the notations for the bulk spacetime. The studied brane embedded in the black hole 
spacetime was codimension-one. The theoretical frameworks of the above-mentioned static and dynamical 
setups, along with computational details and results were presented in sections \ref{sec:sta} 
and \ref{sec:dyn}, respectively. The summary of our findings was presented in section \ref{sec:conclusions}.

\section{Bulk spacetime containing a higher-dimensional dilaton black hole}
\label{sec:bulk}

The considered black hole was a $d$-dimensional dilaton black hole, whose metric in $\left(t,r,\theta_i\right)$ coordinates, where $i=1,...,d-2$, is given by~\cite{gib88}
\ben \label{bh}
ds^2 &=& - \Bigg[ 1 - \bigg( \frac{r_+}{r} \bigg)^{d-3} \Bigg] 
\Bigg[ 1 - \bigg( \frac{r_-}{r} \bigg)^{d-3} \Bigg]^{1 - \ga (d -3)} dt^2 + \nonumber \\
&& + \Bigg[ 1 - \bigg( \frac{r_+}{r} \bigg)^{d-3} \Bigg]^{-1}
\Bigg[ 1 - \bigg( \frac{r_-}{r} \bigg)^{d-3} \Bigg]^{\ga -1} dr^2 
+ r^2 \Bigg[ 1 - \bigg( \frac{r_-}{r} \bigg)^{d-3} \Bigg]^\ga d\Omega_{d-2}^2,
\een
where $r_+$ and $r_-$ are the radii of two horizons, the outer and inner ones, $r_+ > r_-$. The case when
$r_+ = r_-$ is reserved for the extremal higher-dimensional dilaton black hole. 
$\Omega_{d-2}^2$ is the line element of the $d-2$-dimensional unit sphere with coordinates $\theta_i$, $i=1, ..., d-2$, defined on it by the relations $d\Omega_{i+1}^2=d\theta_{i+1}^2+\sin^2\theta_{i+1}d\Omega_i^2$. 
$\ga$ is an auxiliary constant related to the dilaton coupling constant
$\alpha$ and the dimension of the considered spacetime $d$ by the following relation:
\be
\ga = \frac{2 \alpha^2 (d-2)}{(d-3) \left[2 (d-3) + \alpha^2 (d-2) \right]}.
\ee
For the brevity, two functions $f_1 = 1 - \Big( \frac{r_+}{r} \Big)^{d-3}$ and $f_2 = 1 - \Big( \frac{r_-}{r} \Big)^{d-3}$ will be used throughout the paper.

\section{Static brane-dilaton black hole system}
\label{sec:sta}

\subsection{Theoretical setup}
In this section we shall pay attention to a static brane - higher dimensional static spherically symmetric black hole system. A brane $(D-1)$-dimensional configuration in a gravitational field of a black object
will be analyzed having in mind the Dirac-Nambu-Goto action~\cite{dir62,nam70,goto71}
\be
S = \int d^D \zeta \sqrt{-\textrm{det} \ga_{ij}},
\label{action}
\ee
where $\zeta^i$, $i=0, ..., D-1$, are the coordinates on the brane world sheet and the $D$-dimensional metric induced on the world sheet is given by
\ben
\ga_{ij} = g_{\mu\nu} X^\mu_{,i} X^\nu_{,j}
\een
with $X^\mu$, $\mu=0, ..., d-1$, being the bulk spacetime coordinates. 
In the case of a codimension-one brane $D=d-1$. The introduced action~\eqref{action} describes an 
infinitely thin brane. Thus, the analysis of the static configurations of the studied system is 
relevant for such a brane present within it. The problem of thick branes within the geometric construction 
is a subtle one and would require calculations based on perturbative corrections~\cite{czi09,czi10,czi11}.

The assumption that the brane is static and spherically symmetric results in an $O(D-1)$ symmetry of its world sheet. The world sheet is thus defined by the function $\theta_{D-1}\equiv\theta(r)$, with $\theta_i=\frac{\pi}{2}$ for $i=D, ..., d-2$. The metric induced on the brane world sheet has the form
\be
ds^2 = - f_1f_2^{\; 1 - \ga \left( n-3 \right)} dt^2 +
\Bigg[ \frac{f_2^{\; \ga - 1}}{f_1} + r^2 f_2^{\; \ga} \left( \frac{d\theta}{dr} \right)^2 \Bigg] dr^2 +
r^2 f_2^{\; \ga} \sin^2 \theta \ d\Omega^2_n,
\ee
where $n=D-2$. The action \eqref{action} reduces to
\be
S = \Delta T \cA \int dr\; \cL
\ee
with the Lagrange density provided by
\be
\cL = r^n \sin^n\theta \; f_{2}^\frac{\ga n}{2}
\Bigg[ f_{2}^{\; \ga (4-n)} + r^2 f_{1}f_{2}^{\; 1 - \ga(n-4)}
\left( \frac{d\theta}{dr} \right)^2 \Bigg]^\frac{1}{2},
\ee
where $\Delta T$ is the time interval and $\cA$ is the surface of a unit $n$-dimensional
sphere.

Static configurations of the brane-dilaton black hole system are determined by the Lagrange equation
\be
\frac{d}{dr} \bigg( \frac{\p \cL}{\p \theta^\prime} \bigg)
- \frac{\p \cL}{\p \theta} = 0,
\ee
where $^\prime$ denotes the differentiation with respect to the $r$-coordinate. An explicit form
of the equation is provided by
\be
\label{brane_static}
\frac{d^2 \theta}{d r^2} + A_1\left( \frac{d \theta}{d r} \right)^3 +
A_2\left( \frac{d \theta}{d r} \right)^2 + A_3\left( \frac{d \theta}{d r} \right) + A_4 = 0,
\ee
where by $A_i,~i=1,\dots, 4$, we have denoted the following:
\ben
\label{s_A1}
A_1 &=& r(n+1) f_1 f_2 + \frac{r^2}{f_2^{\; \gamma (4-n) + \frac{\gamma n}{2}}}~\frac{d}{dr} \bigg(
f_1 f_2^{\; 1 - \gamma (n-4) +\frac{\gamma n}{2}} \bigg)  + \nonumber \\ 
&&- \frac{1}{2}~\frac{r^2}{f_2^{\; \gamma (4-n)}}~\frac{d}{dr}
\bigg(f_1 f_2^{\; 1 - \gamma (n-4) } \bigg) = \nonumber \\
&=& r(n+1) f_1 f_2 + \frac{r^2}{2} f_2 \frac{d f_1}{ d r} + \frac{r^2}{2} (1 + 4 \gamma ) f_1 \frac{d f_2}{d r}, \\
\label{s_A2}
A_2 &=& -n \cot\theta, \\
\label{s_A3}
A_3 &=& \frac{n+2}{r} + \frac{1}{f_1 f_2^{\; 1- \frac{\gamma n}{2} + 4 \gamma}}~\frac{d}{dr} \bigg(
f_1 f_2^{\; 1- \frac{\gamma n}{2} + 4 \gamma} \bigg) - \frac{1}{2~f_2^{\; \gamma (4-n)}}~\frac{d}{dr} \bigg( f_2^{\; \gamma (4-n)} \bigg) = \nonumber \\
&=& \frac{n+2}{r} +  \frac{1}{f_1}\frac{d f_1}{d r} + (1 + 2 \gamma )\frac{1}{f_2}\frac{d f_2}{d r}, \\
\label{s_A4}
A_4 &=& - \frac{n \cot\theta}{r^2 f_1 f_2}.
\een

The equation \eqref{brane_static} possesses singular points at the black hole horizon and at infinity. In order to derive static configurations of the considered system, the behavior of the 
brane in the two limiting regions, i.e., $r \rightarrow r_{+}$ and $r \rightarrow \infty$, has to be determined. This behavior gives appropriate boundary conditions for the $\theta$ function, which differ 
for nonextremal and extremal black holes. The boundary behavior will be discussed for the nonextremal BDBH system and then the differences relevant to the extremal case will be pointed out.

\subsection{Brane behavior in the limiting regions}
In principle,  there exist two possible configurations of the brane in the near horizon limit $r \rightarrow r_{+}$, which are the brane penetrating and not touching the black hole event horizon. The analysis of the penetrating 
solution reveals the singular coefficients in the equation (\ref{brane_static}) which are composed of the part of the second expression in (\ref{s_A3}) and the $A_4$ coefficient. 
Hence, the singular part of the considered equation is of the form
\ben
\widetilde{A}_3\: \theta^\prime + A_4,
\een 
where $\widetilde{A}_3 = f_1^{\; -1} f_1^\prime$. It can be written as
\ben
\frac{f_1^\prime \theta^\prime - n \frac{\cot\theta}{r^2 f_2}}{f_1} \sim \frac{0}{0},
\een  
where the right-hand side of the above relation is connected with the regularity demand. The above expression yields
\ben
\left(f_1^\prime \theta^\prime - n \frac{\cot\theta}{r^2 f_2} \right) \bigg|_{r = r_{+}} = 0.
\een
Postulating that $\theta|_{r \rightarrow r_{+}} \approx \theta_0  + \theta_1 (r - r_{+})$, we get
that $\theta_0$ is an arbitrary constant and 
\ben
\theta_1 = \frac{n \cot{\theta_0}}{f_1^\prime r^2 f_2} \bigg|_{r = r_{+}}.
\een
The above analysis gives both the form of the boundary condition on the horizon $\theta = const.$ and 
the near horizon behavior of $\theta$ which is helpful in a numerical integration of the discussed equation.

In the case when the brane does not touch the black hole horizon, the input parameter dictating 
the boundary condition is the minimal distance between the brane and the origin of the coordinate system, $r_0$. 
Due to the symmetry of the system, this distance should be attained for $\theta = 0$, which again leads to the singularity of 
the equation (\ref{brane_static}) in the $\cot\theta$ term. Postulating that $\theta|_{r \rightarrow r_{0}} \approx \eta_0 (r - r_0)^{\frac{1}{2}} + \eta_1 (r - r_0)^{\frac{3}{2}}$, plugging it into 
the relation \eqref{brane_static} and expanding the $\theta$-dependent term around $r = r_0$, the following relations for the $\eta_0$ and $\eta_1$ coefficients arise:
\ben
\eta_0 &=& \pm \sqrt{\frac{2}{A_1} (1 - a_2 )} \bigg|_{r = r_0}, \\
\eta_1 &=& \left. \pm \frac{2 \sqrt{2 \left( 1 - a_2 \right)} \Big[ a_2 - 2 a_2^2 + a_2^3 + 3A_1 a_2 A_3 - 3 A_1 \left( A_3 + A_1 a_4 \right) \Big]}
{ 9 A_1^{\frac{3}{2}} \left( 3 - 4 a_2  + a_2^2 \right)} \right|_{ r = r_0},
\een 
where $a_2$ and $a_4$ are the $\theta$-independent parts of $A_2$ and $A_4$, respectively.

Having specified the behavior of the investigated function in the near horizon limit, we will turn to the $r \rightarrow \infty$ one. Assuming that the brane behaves in this limit as a $d-1$-dimensional plane we may write
\ben
\theta = \frac{\pi}{2} + \lambda(r).
\een
Moreover, it is assumed that the solution approaches $\theta = \frac{\pi}{2}$ at least as fast as $\frac{1}{r}$. 
An analysis of the coefficients of the considered equation \eqref{s_A1}--\eqref{s_A4} leads to the relations
\ben\label{eqn:A11}
A_1 &\sim& r^2, \\
A_2 &\sim&  n \lambda(r), \\
A_3 &\sim& \frac{n+2}{r} - (n - 1) r_+^{\; n-1} r^{-n} - (n-1) (1 + 2 \gamma )r_-^{\; n-1} r^{-n}, \\
A_4 &\sim& \frac{n}{r^2} \lambda(r).
\label{eqn:A44}
\een
Since $\lambda$ decays like $\frac{1}{r}$ or faster by definition, the equation \eqref{brane_static} becomes
\ben
\label{brane_asym}
\lambda^{\prime\prime} + \bigg[\frac{n+2}{r} - (n-1) r_+^{\; n-1} r^{-n} - (n-1) (1 + 2 \gamma) r_-^{\;n-1} r^{-n} \bigg] \lambda^\prime + \frac{n}{r^2} \lambda = 0. 
\een  
The second and third terms in the square bracket are relevant only for $d=3$, as for the bigger number of 
dimensions they decay as $r^{-2}$ or faster and thus are irrelevant. This implies that for $d \geqslant 4$ a charged brane behaves in the limit of $r\to\infty$
like an uncharged one. In this case the solution to the equation (\ref{brane_asym}) is given by
\ben
\lambda = c_1 r^{-1} + c_2 r^{- n},
\een
where $c_1$ and $c_2$ are integration constants. 
In the case of $d=3$ we have $n=1$ and despite the fact that the terms $r^{-n}$ are relevant, the coefficients in front of them vanish. 
So the solution is again the same as in the uncharged case and is provided by~\cite{fro06}
\ben
\lambda = c_1 r^{-1} + c_2 \frac{\ln r}{r}.
\een

After investigation of the behavior of the coefficients $A_i$ of the equation \eqref{brane_static} we conclude that in the discussed limit of $r\to\infty$ they differ very little from the uncharged brane case, which is evident from the relations \eqref{eqn:A11}-\eqref{eqn:A44}. Moreover, this difference is actually irrelevant in the large $r$ limit. Basing on this, we assume that the decay rate should be polynomial. Inserting $\lambda \sim r^{-\alpha}$ into the equation, where $\alpha$ is a positive number, we obtain the following conclusions. 
For any $\alpha >0$ the term $A_2 \lambda^{' 2}$ is always subleading. For $\alpha \geqslant 1$ we have the case discussed above, and on the other hand a closer inspection reveales that for $\alpha > \frac{1}{2}$ the term $A_1 \lambda^{' 3}$ is also always subleading. Stating this, the only remaining possibility is $\alpha\in\langle\frac{1}{2},0)$, as in this case the $A_1 \lambda^{'3}$ term becomes relevant and the equation becomes nonlinear. Moreover, in this case we were unable to find a solution in the form of the known special functions. The same is true also for the uncharged brane discussed in~\cite{fro06}.

When an extremal dilaton black hole is present in the BDBH system, the behavior of the brane at infinity and near the horizon when the two coexisting objects are separated is the same as discussed above for the nonextremal case. 
On the other hand, for the brane touching the degenerate horizon the regularity condition implies
\ben
\frac{\Big[ f_2 f_1^\prime + ( 1 + 2 \gamma ) f_1 f_2^\prime \Big] \theta^\prime - \frac{n \cot\theta }{r^2}}{f_1 f_2} \sim \frac{0}{0}.
\een 
It was derived analogously to the nonextremal case and it implies that $\cot\theta = 0$. Together with an assumption $\theta(r) \approx \theta_0 + \theta_1 (r-r_+)$ it gives that $\theta_0 = \frac{\pi}{2}$.
Using the L'Hospital rule once results in the relation
\ben
\frac{\Big[ f_2 f_1^\prime + \left( 1 + 2 \gamma \right) f_1 f_2^\prime \Big] \theta^{\prime\prime}}
{f_1^\prime f_2 + f_1 f_2^\prime} &+& \\ \nonumber 
+ \frac{\Big[ f_2^\prime f_1^\prime + f_2 f_1^{\prime\prime} + \left( 1 + 2 \gamma \right) f_1^\prime f_2^\prime + 
\left( 1 + 2 \gamma \right) f_1 f_2^{\prime\prime} \Big] \theta^\prime
+ \frac{2 n \cot\theta}{r^3} + \frac{n}{r^2 \sin^2\theta} \theta^\prime}
{f_1^\prime f_2 + f_1 f_2^\prime}  &\sim& \frac{0}{0}. 
\een  
Since $f_1(r_{+}) = f_2(r_{+}) = 0$, the coefficient in the front of the second derivative of $\theta$, trivially vanishes. 
Vanishing of the nominator of the second component yields
\ben
\theta_1 = - \frac{2 n \cot\theta_0}{r_{+}^2} \left[
\frac{n}{r_{+}^2 \sin^2\theta_0} + 2 \left ( 1 + \gamma \right ) f_1^\prime f_2^\prime
\right]^{-1},
\een
where the expansion $\theta|_{r \rightarrow r_{+}} \approx \theta_0 + \theta_1 (r-r_+)$ was employed. 
The trigonometric functions of $\theta$ were also expanded and only the leading terms were kept. Since $\theta_0$ is already fixed to be equal to $\frac{\pi}{2}$, we conclude that $\theta_1=0$. Hence, in the 
extremal case the brane will either not touch the black hole horizon or will touch it precisely at one point on the equator. Moreover, in the last case the brane configuration is given by $\theta = \frac{\pi}{2}$.
Due to the near horizon behavior of the brane this is true for any spacetime dimension $d$ and an arbitrary dilatonic coupling constant $\alpha$.

\subsection{Spatial configurations}
\label{ssec:spaconf}

The static spatial arrangements of the BDBH system were investigated for the bulk spacetime dimensions $5$, $6$, $7$ and $8$ and the dilatonic coupling constant equal to $0$, $-1$ and $-\sqrt{3}$. The 
studied values of $\alpha$-coupling constant referred to the uncoupled dilaton field, the low energy limit of the string theory and the dimensionally reduced Kaluza-Klein theory, respectively. Each brane configuration in the black hole spacetime will be depicted as a function $\theta\left(r\right)$ via cylindrical coordinates $Z=r\cos\theta$, $R=r\sin\theta$.

We would like to stress that the numerical solutions of the full equation \eqref{brane_static} were obtained without any assumptions on the behavior of the solution in the limit $r\to\infty$. In other words, integrating this equation from the black hole horizon to very large values of the $r$ coordinate ($r_{max} = 10^4 r_{+}$), we found out that it indeed approaches $\frac{\pi}{2}$.

\begin{figure}[h]
\centering
\includegraphics[width=0.475\textwidth]{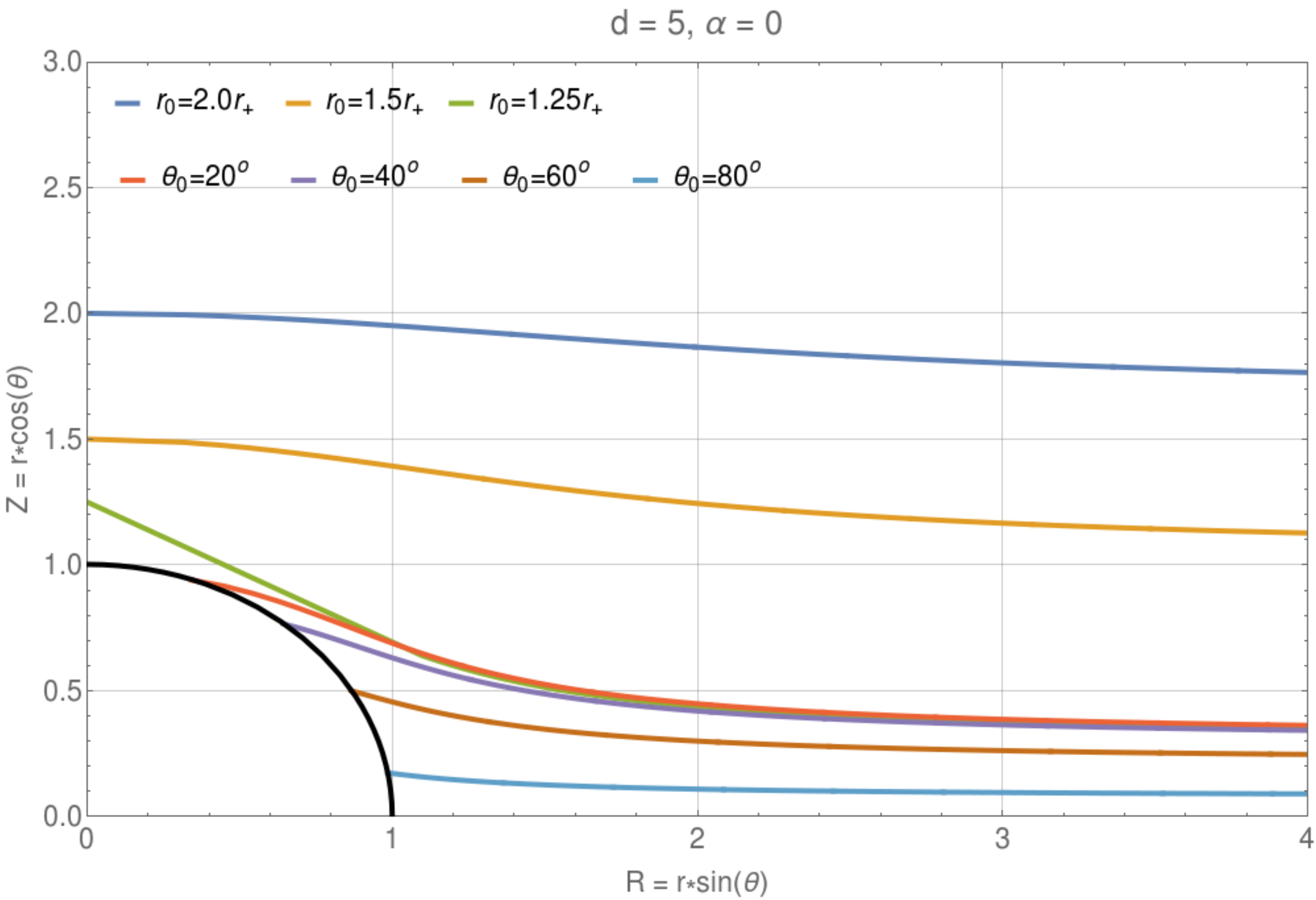}
\includegraphics[width=0.475\textwidth]{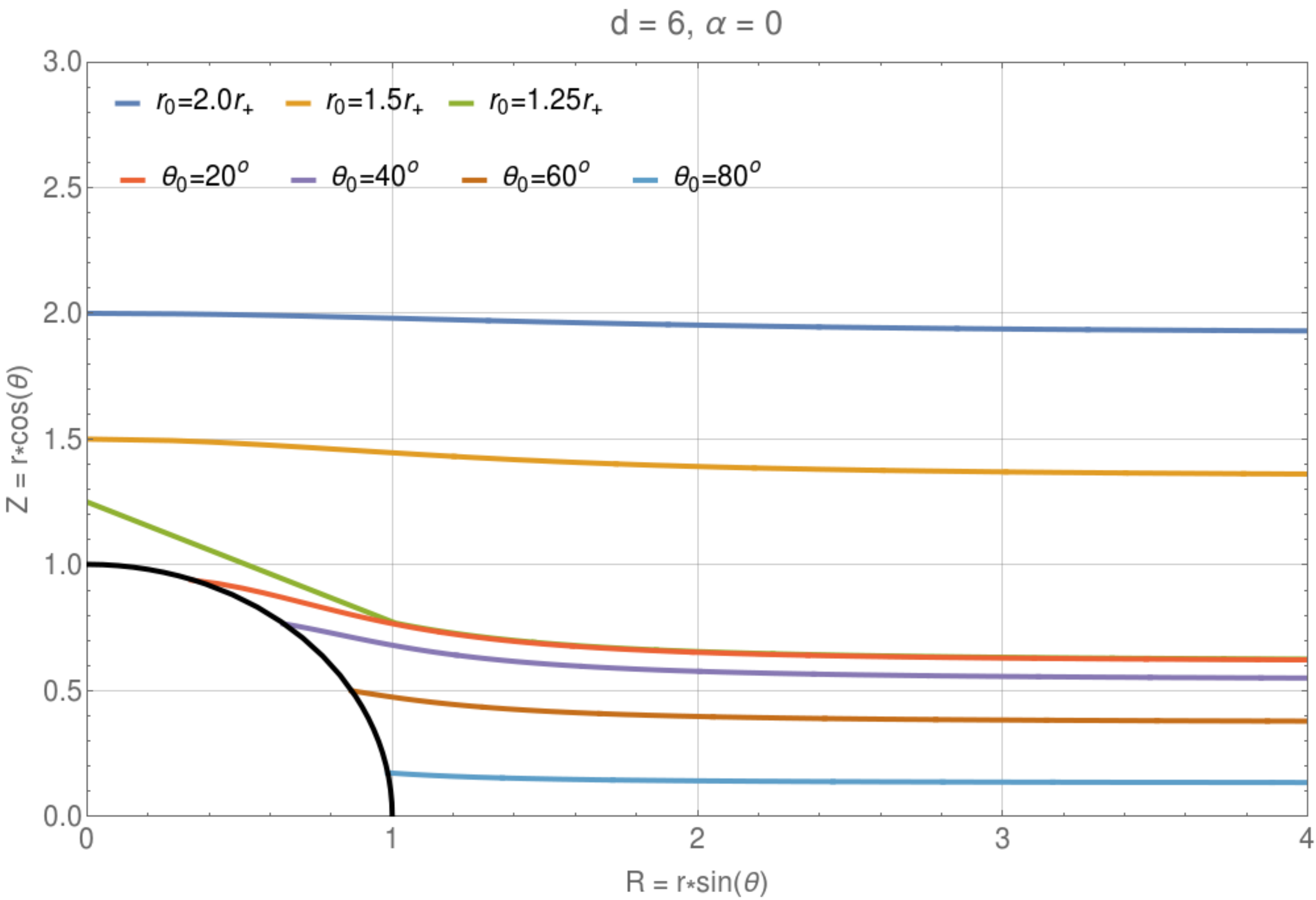}
\vspace{0.15cm}\\
\includegraphics[width=0.475\textwidth]{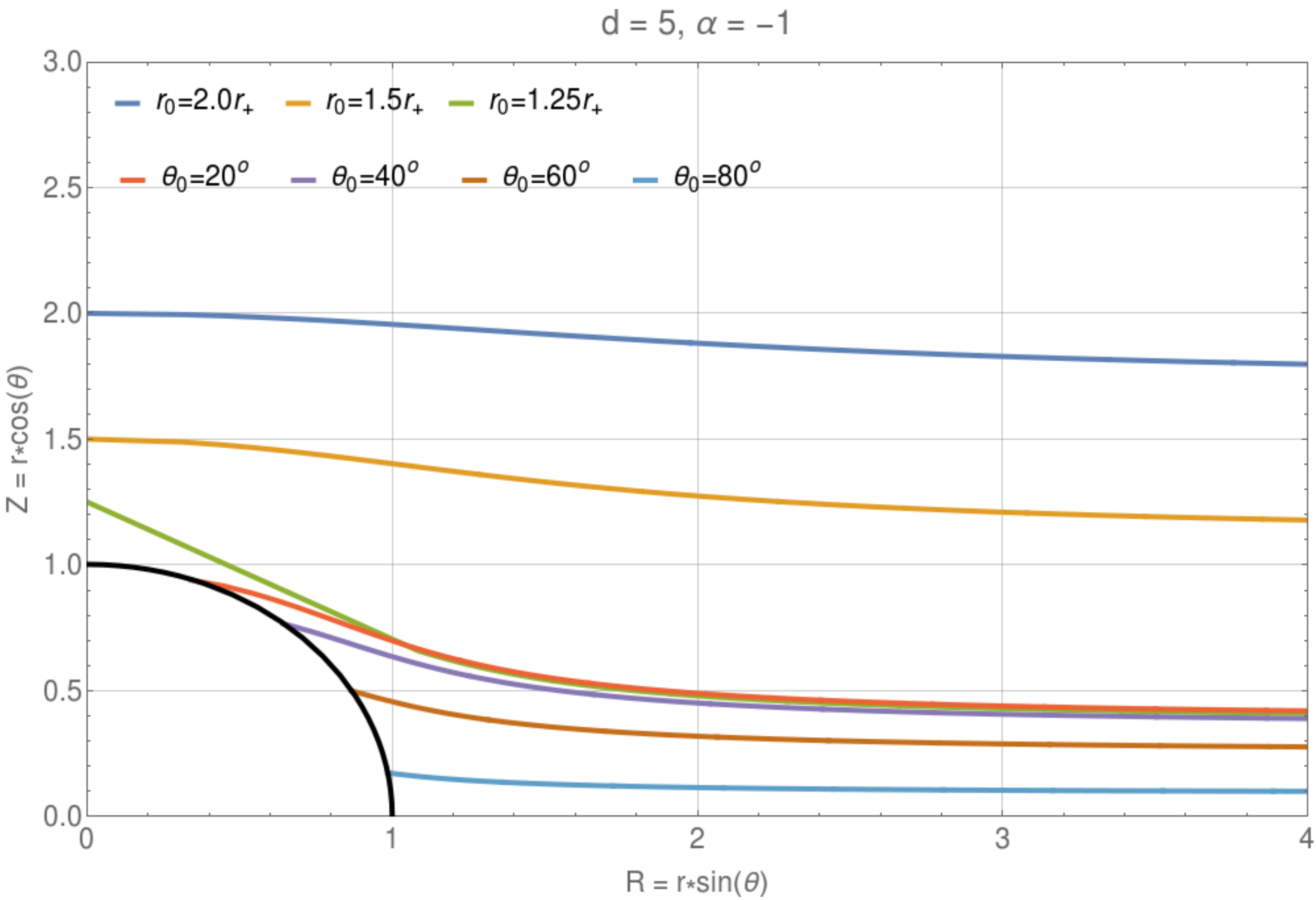}
\includegraphics[width=0.475\textwidth]{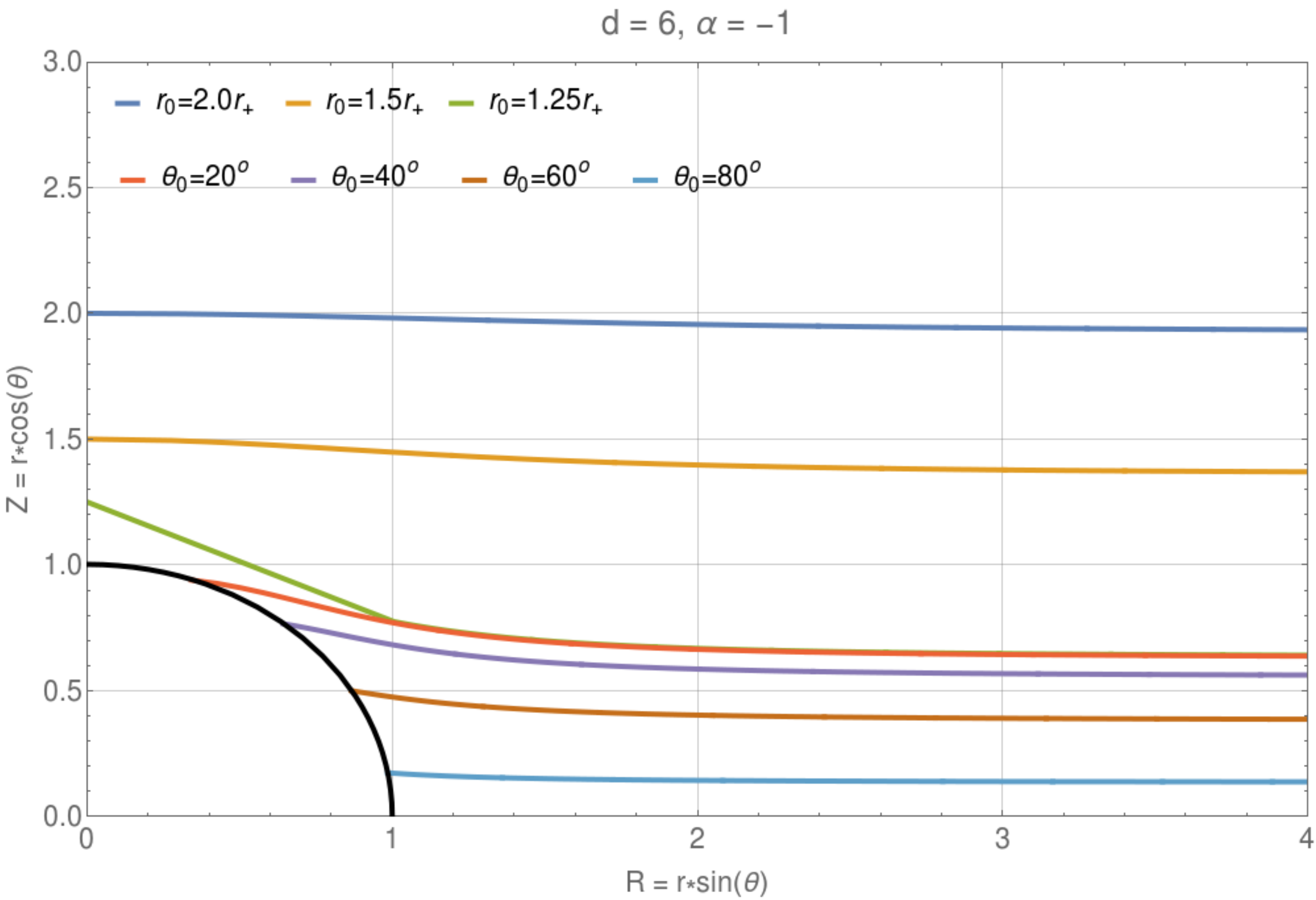}
\vspace{0.15cm}\\
\includegraphics[width=0.475\textwidth]{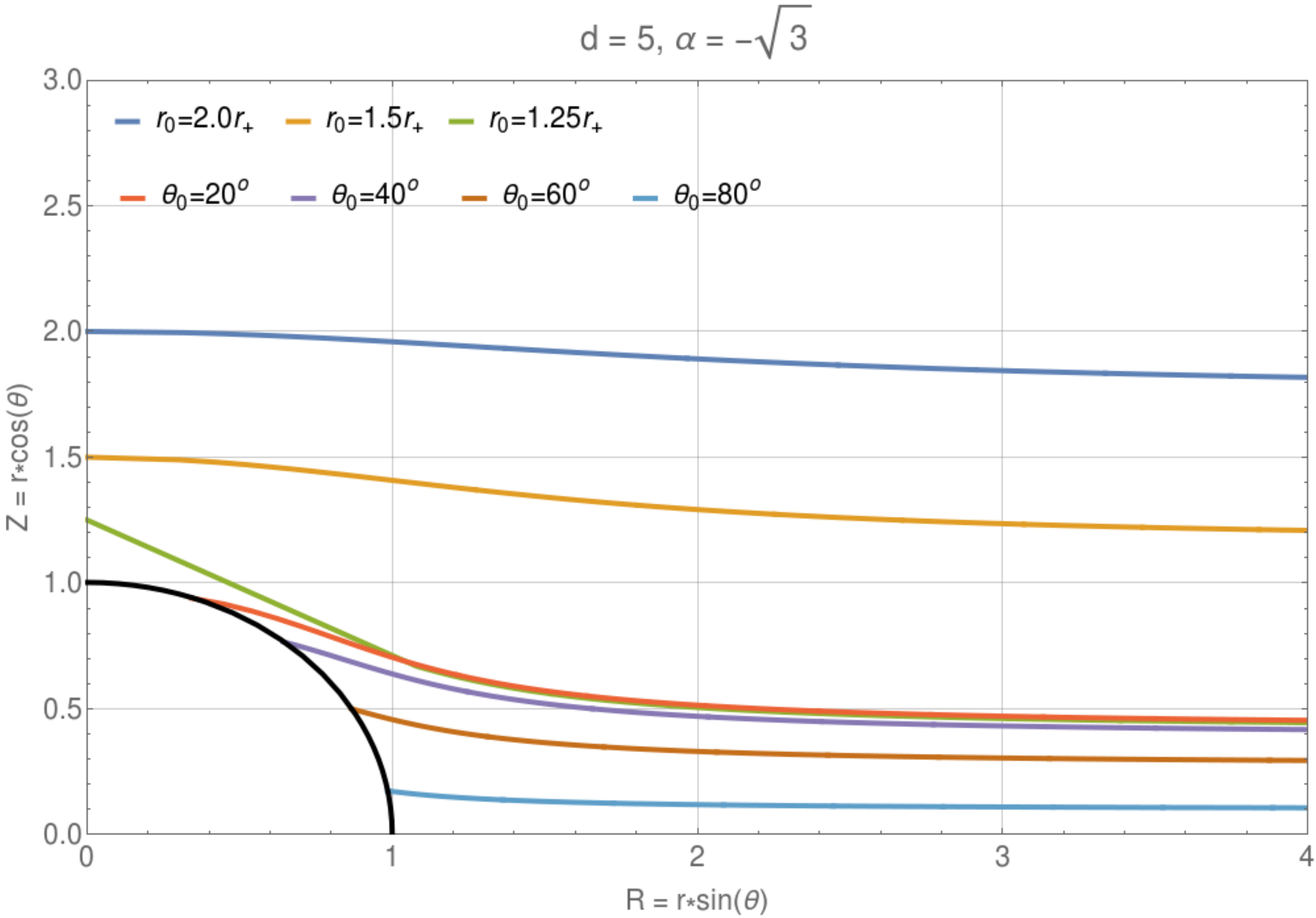}
\includegraphics[width=0.475\textwidth]{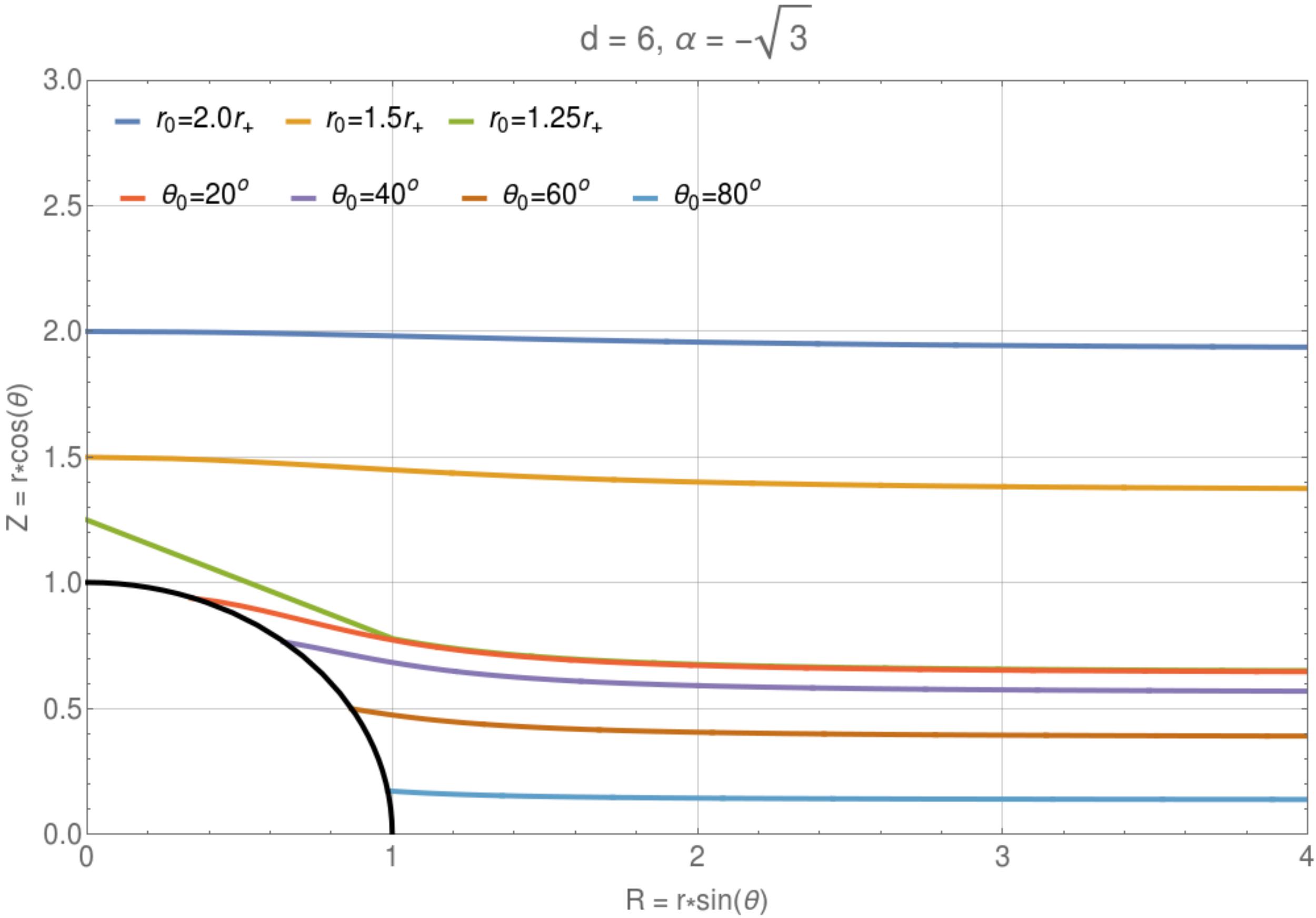}
\caption{Static configurations of the BDBH system with a nonextremal black hole for $d$ equal to $5$ and $6$. The spacetime dimension and the value of the coupling constant $\alpha$ are presented above each diagram. The parameters $r_0$ and $\theta_0$ corresponding to the particular brane locations are listed on the plots.}
\label{fig:sta-nonextr-1}
\end{figure}

\begin{figure}[h]
\centering
\includegraphics[width=0.475\textwidth]{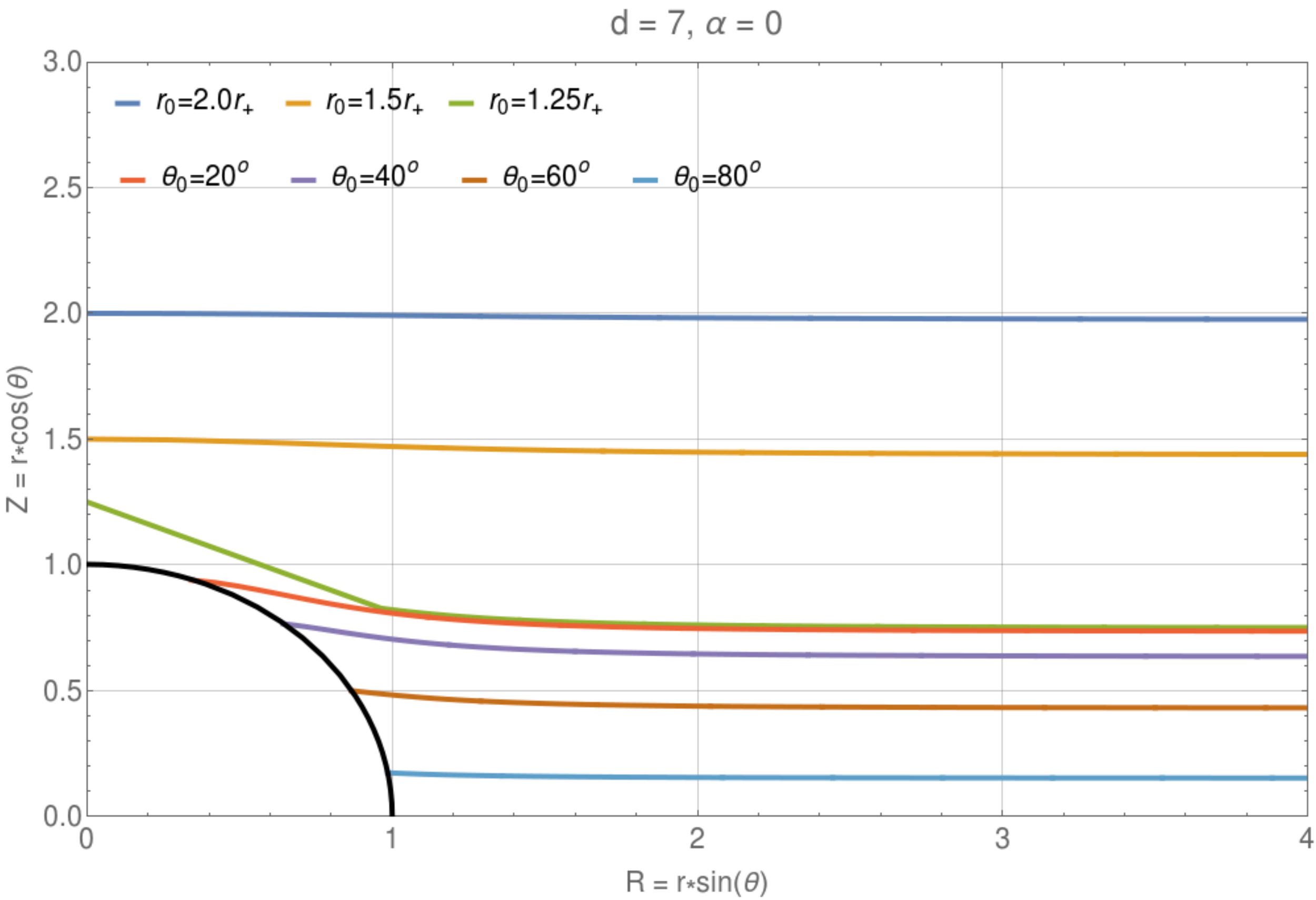}
\includegraphics[width=0.475\textwidth]{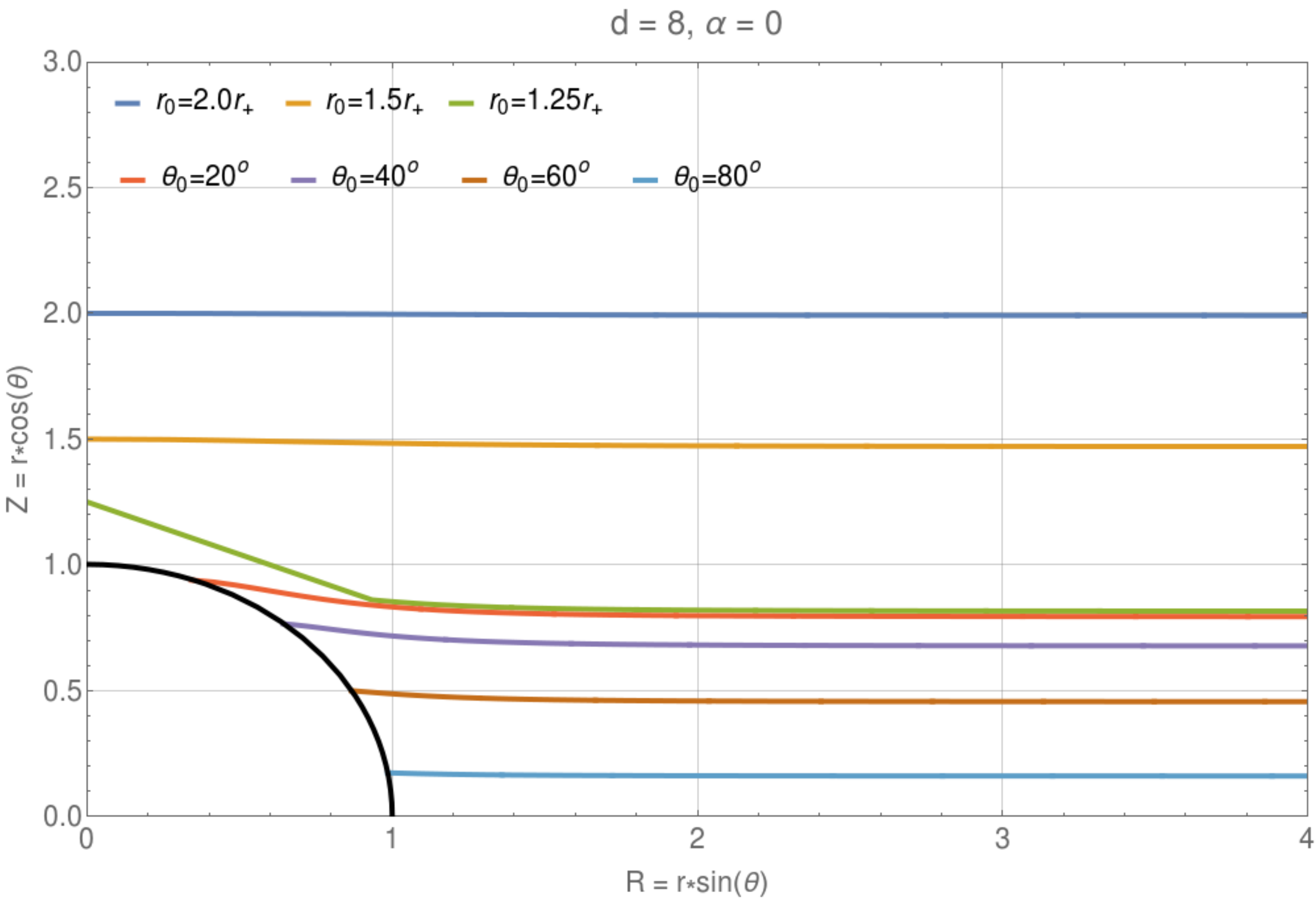}
\vspace{0.15cm}\\
\includegraphics[width=0.475\textwidth]{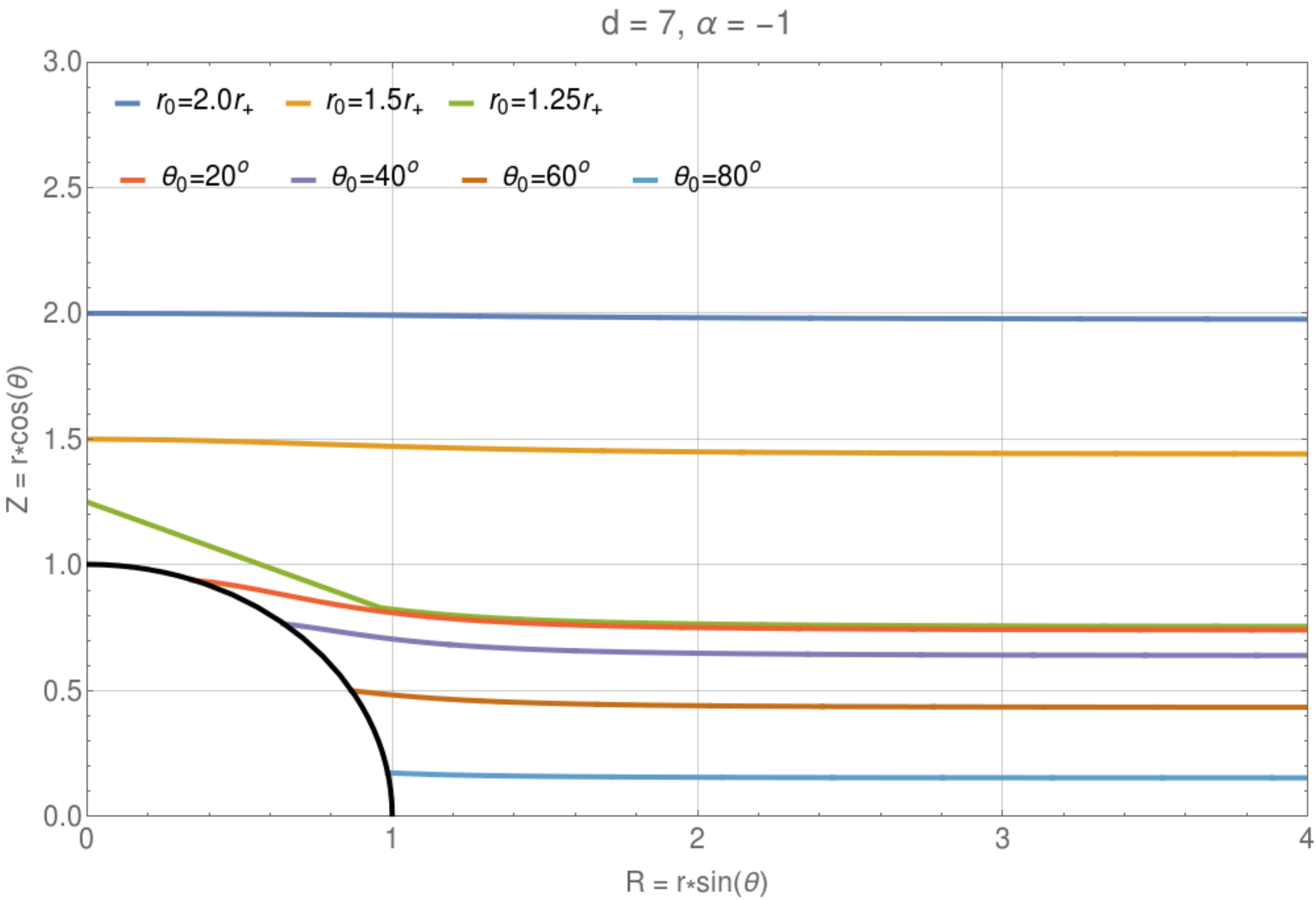}
\includegraphics[width=0.475\textwidth]{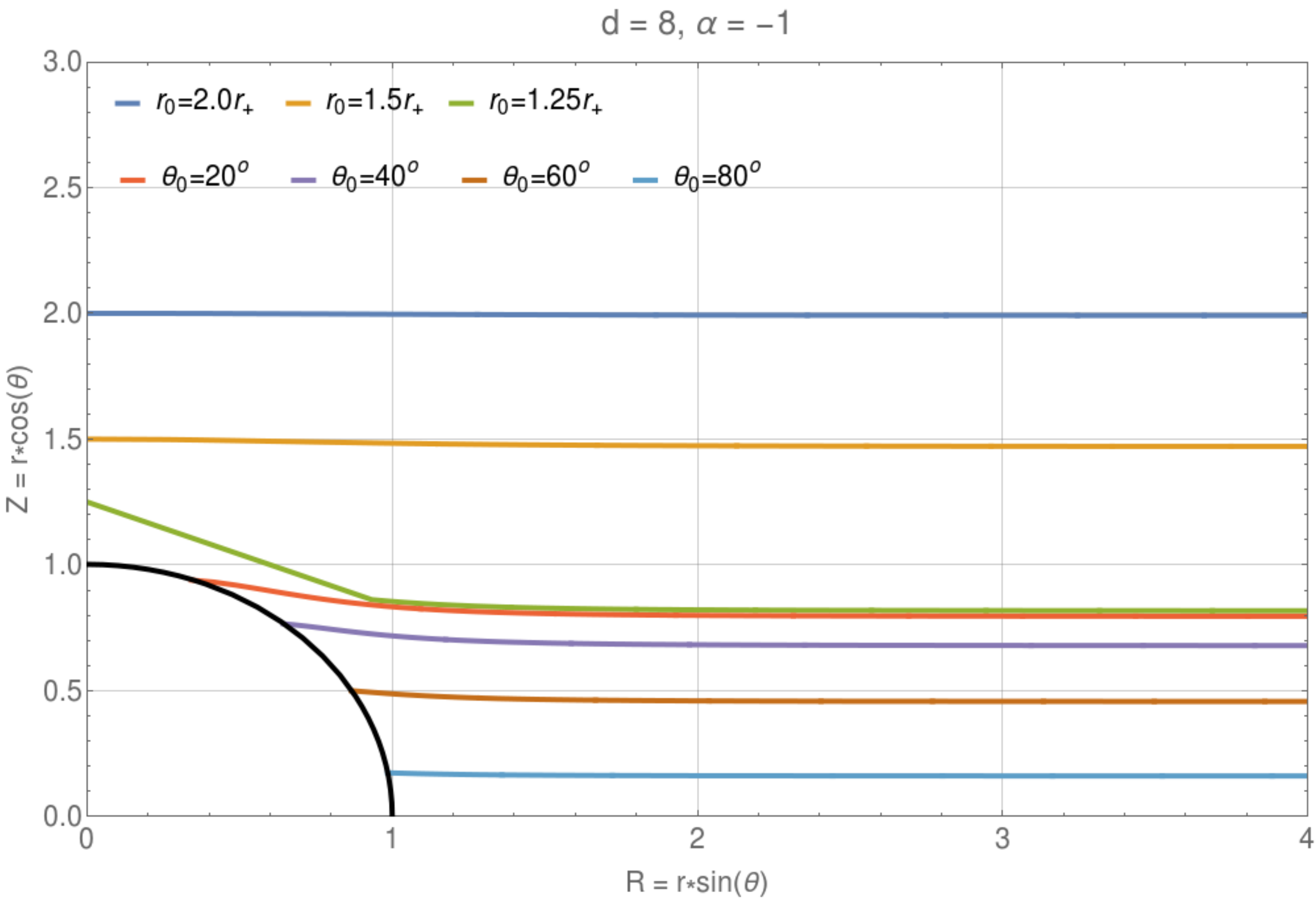}
\vspace{0.15cm}\\
\includegraphics[width=0.475\textwidth]{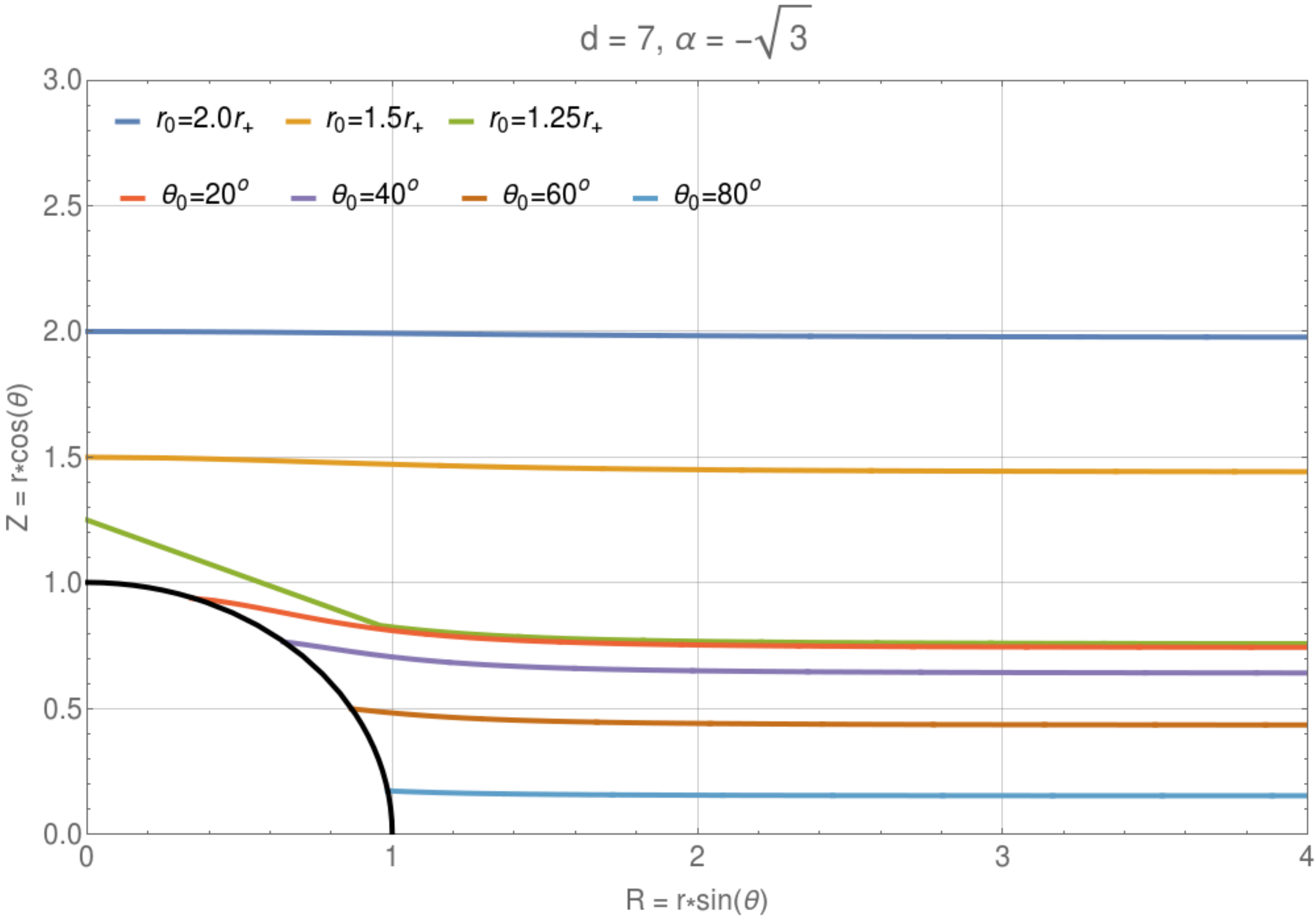}
\includegraphics[width=0.475\textwidth]{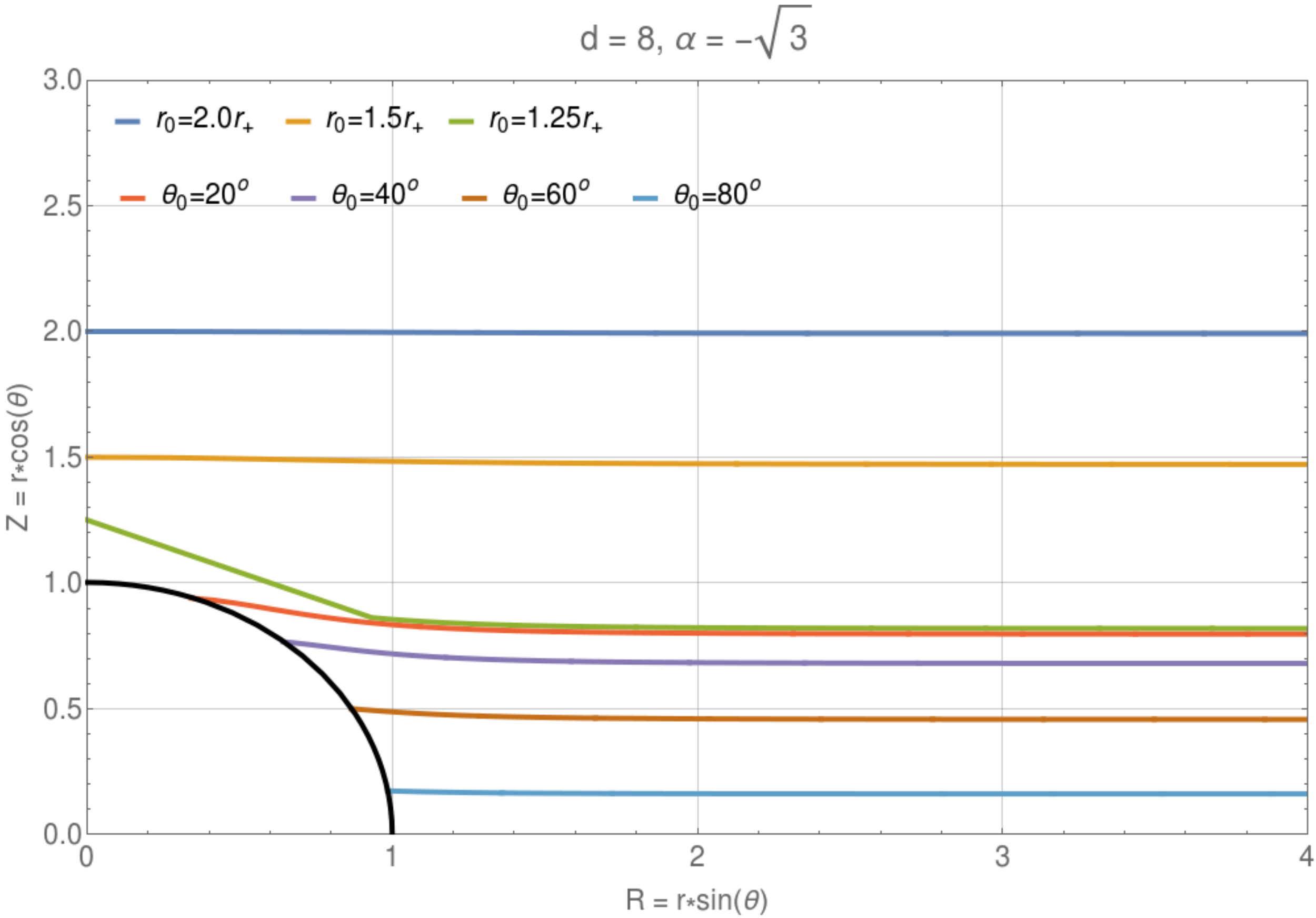}
\caption{Static configurations of the BDBH system with a nonextremal black hole for $d$ equal to $7$ and $8$. The spacetime dimension and the value of the coupling constant $\alpha$ are presented above each diagram. The parameters $r_0$ and $\theta_0$ corresponding to the particular brane locations are listed on the plots.}
\label{fig:sta-nonextr-2}
\end{figure}

The configurations of the static nonextremal brane-dilaton black hole system are presented in figures \ref{fig:sta-nonextr-1} and \ref{fig:sta-nonextr-2}. The nonextremal black hole admits a brane which intersects its horizon and hence both near horizon configurations mentioned in the previous section are observed in this case. The non-penetrating configurations correspond to various values of the parameter $r_0$, and the intersecting branes are related to different values of $\theta_0$. The value of the dilatonic coupling constant does not impact the static configurations. The number of spacetime dimensions influences the brane location in the black hole spacetime such that the brane bend in the vicinity of the black hole horizon disappears closer to it as $d$ increases. 
It will be interesting to propose the physical interpretation of the observed property. Because of the fact that gravity can penetrate the additional dimensions, its strength is weaker
in the case of the growth of the dimensionality of the underlying spacetime, comparing to the gravity force exerted by black hole in four-dimensions. Therefore the increase of the spacetime dimensions (decrease
of the gravity force which should penetrate the additional dimensions) is responsible for the closer to the event horizon disappearance of the brane. The very similar situation was
revealed in \cite{rog08br,rog09br,gib08br}, where it was found that the absorption probability of massive Dirac fermions in the spacetime of higher-dimensional Schwarzschild black hole and a tense brane black hole, decreased
with the increase of the spacetime dimensionality.

\begin{figure}[h]
\centering
\includegraphics[width=0.475\textwidth]{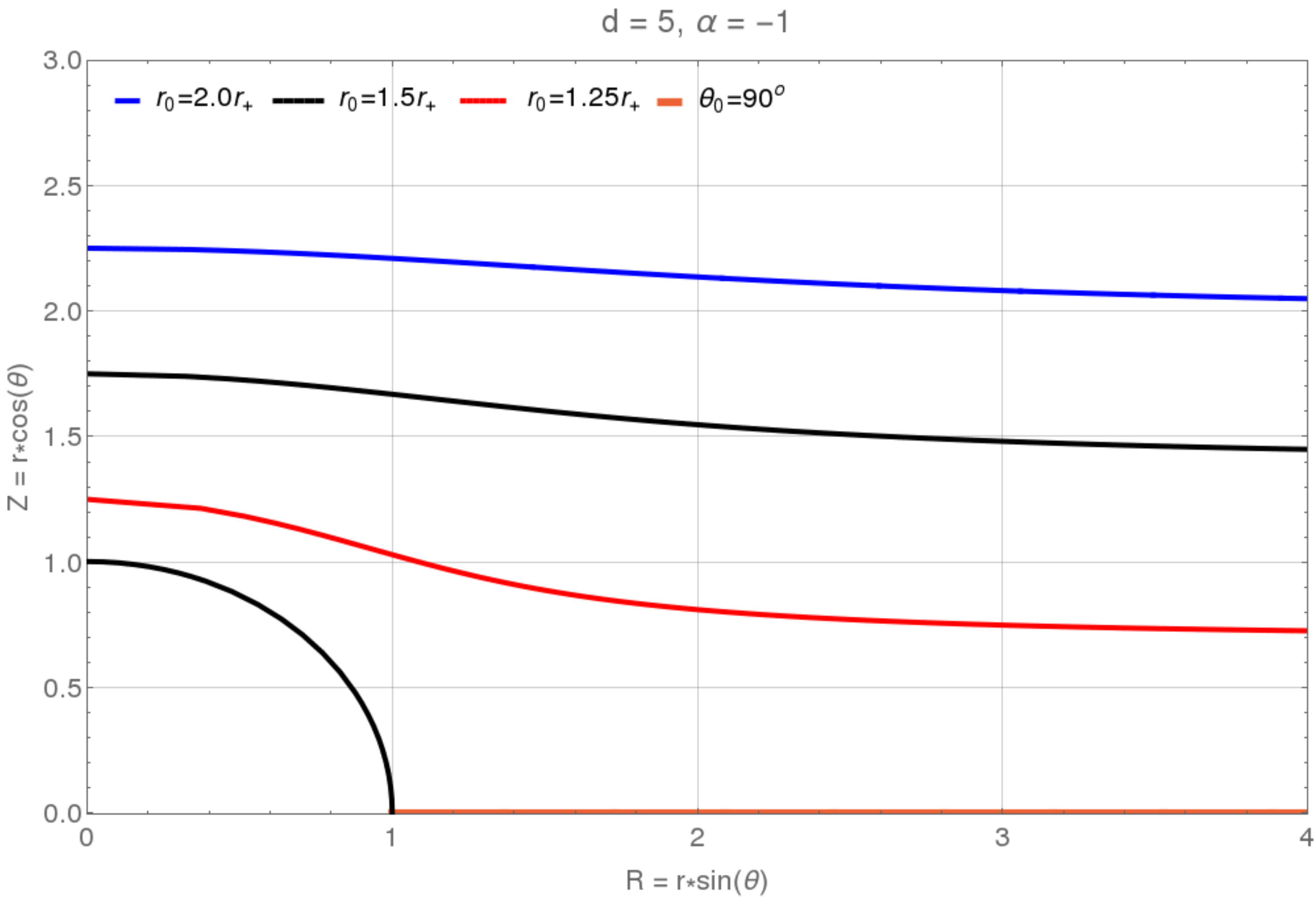}
\includegraphics[width=0.475\textwidth]{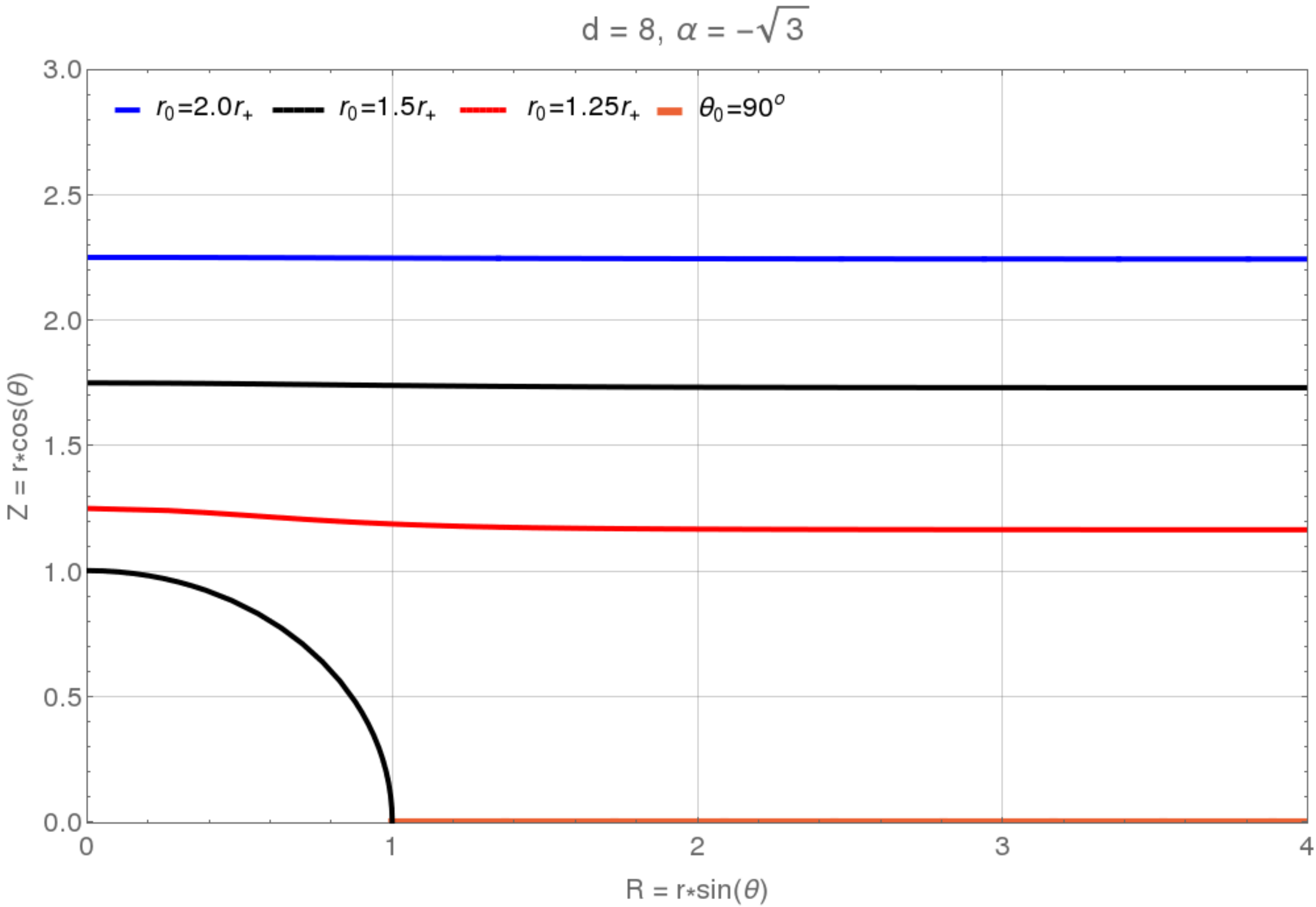}
\caption{Static configurations of the BDBH system with an extremal black hole. The spacetime dimension $d$ and the value of the coupling constant $\alpha$ are presented above each diagram. The parameters $r_0$ and $\theta_0$ corresponding to the particular brane locations are listed on the plots.}
\label{fig:sta-extr}
\end{figure}

The static configurations of an extremal brane-dilaton black hole system are depicted in figure \ref{fig:sta-extr}. Disregarding the values of $d$ and $\alpha$, there exists only one brane location which penetrates the horizon and it corresponds to the brane situated on the equator plane of the black hole, with $\theta_0=\frac{\pi}{2}$ and hence $\theta=\frac{\pi}{2}$. The remaining arrangements represent branes which do not have any common points with the horizon. Each of these branes is described by a different value of $r_0$. Similarly to the nonextremal case, the bend of the brane disconnected with a black hole becomes smaller nearby its event horizon as the dimensionality of
the considered spacetime $d$ increases. 
The above facts are in accord with the conclusions gained in four-dimensions, where the dilaton
black hole - domain wall (brane) system was studied \cite{mod03}. Using the scalar field with various
potentials, i.e., $\phi^4$ and sine-Gordon ones, the domain wall behavior in the spacetime
of charged dilaton black hole was simulated numerically. The conclusion was that for the extremal
charged dilaton black hole one always obtained the expulsion of the domain wall.

\section{Evolution of the brane-dilaton black hole system}
\label{sec:dyn}

Our further investigations concentrated on the dynamics of the BDBH system. The tension of the domain wall was assumed to be small which allowed us to ignore its self-gravitation effect.

\subsection{Theoretical model}
\label{ssec:dyn-model}

When describing the brane, we confined our attention to a scalar effective field theory with a quadratic potential 
of the scalar field $\Phi$ given by the relation
\be
V \left( \Phi \right) = \frac{\la}{4} \Big( \Phi^2 - \eta^2 \Big)^2,
\ee
where the parameters are related to the tension of a domain wall solution in a flat space $\sigma\approx\sqrt{\lambda}\eta^3$ and its thickness $l\approx(\sqrt{\lambda}\eta)^{-1}$.
In contrast to the analysis of the static configurations of the examined system conducted in the 
preceding section, the effective field theory approach employed in investigations of the dynamical 
case allowed us to study not only an infinitely thin brane, but also branes of bigger thicknesses.

The equation of motion of the domain wall intersecting the dilaton black hole completely immersed in the
higher-dimensional spacetime yields
\be
\na_\mu \na^\mu \Phi - \frac{\p V}{\p \Phi} = 0.
\ee
If one assumes an $O(d-2)$-rotational symmetry around the axis perpendicular to the equatorial plane of the considered black hole, then the following explicit form of the equation of motion arises:
\be
\frac{-1}{f_1 f_2^{\; 1 - \ga (d-3)}}\; \p_t^2 \Phi +
\frac{1}{r^{d-2} f_2^{\; \ga}}\ \p_r \Big( r^{d-2} f_1 f_2 \p_r \Phi \Big)
+ \frac{1}{r^2 \left( \sin\theta \right)^{d-3}}\; \p_\theta \Big[ \left( \sin\theta \right)^{d-3} \p_\theta \Phi \Big] 
- \frac{\p V}{\p \Phi} = 0.
\label{eqn:brane}
\ee

Our main aim was to examine the potential recoil of the dilaton black hole from the brane during a dynamical process, at the beginning of which the brane intersecting 
the black hole acquires an initial velocity. Therefore, $\Phi$ and $\p_t \Phi$ for $t=0$ are given by a static kink
profile boosted in the direction of the symmetry axis~\cite{fla06a}. Namely, one has
\be
\Phi_v = \eta \tanh \Bigg( \sqrt\frac{\la}{2} \eta \frac{r \cos\theta - vt}{\sqrt{1 - v^2}} \Bigg),
\label{boo}
\ee
where $v$ is constant.
\par
At infinity, the field describing the brane reduces to the case of a flat spacetime form given by the relation \eqref{boo},
while the regularity conditions at the symmetry axis and on the black hole event horizon are
\ben
\frac{\p \Phi}{\p \theta} &=& 0 \quad \textrm{at } \theta = 0, \pi, \\ \nonumber
\frac{\p \Phi}{\p t} &=& 0 \quad \textrm{at } r = r_+.
\een

\subsection{Numerical computations}
\label{ssec:dyn-computations}

The equation of motion of the scalar field representing the brane \eqref{eqn:brane} 
was solved using the collocation spectral method~\cite{Boyd} in the spatial directions $r$ and $\theta$. 
The temporal evolution was conducted via a finite difference scheme based on three adjacent $t=const.$ hypersurfaces.

The applied pseudospectral method required writing the equation to be solved using a polynomial expansion of an unknown function at the so-called collocation points. The points were spread on a grid, which spanned from $-1$ to $1$ in the spectral directions, which in the considered case were $r$ and $\theta$.

The general form of $\Phi$ was chosen as an expansion in terms of the Chebyshev polynomials of the first kind
\ben
\Phi \big( t,y(r),x(\theta) \big) &=& \sum\limits_{m=0}^M \sum\limits_{n=0}^N \varphi_{mn}(t)\ T_m(y)\ T_n(x),
\label{eqn:exp}
\een
where $y=\frac{r-2r_+-1}{r+1}$ and $x=\cos\theta$ are the independent spatial variables rescaled to the range $[-1,1]$, 
which is necessary for applying the pseudospectral method. The time dependent expansion coefficients were denoted as $\varphi_{mn}$. The applied expansion involved $50$ Chebyshev polynomials in each of the two spatial directions. 
The collocation points were of the Gauss-Lobatto type, i.e, they were spread on the computational grid, in each of the spectral directions, according to a prescription
\ben
z_k &=& -\cos \frac{k\pi}{K},
\label{eqn:coll}
\een
where $k=0,\dots,K$, $z$ stands for either $y$ or $x$ and $K$ denotes $M$ or $N$, respectively.

The equation of motion of $\Phi$  \eqref{eqn:brane} was reduced by applying the expansion \eqref{eqn:exp} to the evolution equation of the expansion coefficients $\varphi_{mn}$
and obtained the form
\ben
\sum\limits_m \sum\limits_n \ddot{\varphi}_{mn}(t)\ T_m(y)\ T_n(x) 
- \frac{(1-y)^d f_1 f_2^{\; 1 - \ga (d-2)}}{2(r_++1)(2r_++1+y)^{d-2}} 
\sum\limits_m \sum\limits_n \varphi_{mn}(t)\ T_n(x)\ L_m(y) + \nonumber\\
- \frac{(1-y)^2~f_1 f_2^{1 - \gamma (d-3)}}{(2r_++1+y)^2} \sum\limits_m \sum\limits_n \varphi_{mn}(t)\ T_m(y)\ L_n(x) + f_1 f_2^{1 - \gamma (d-3)}~\frac{dV}{d\Phi} = 0,
\hspace{0.75cm}
\label{eqn:ev-coeff}
\een
where a dot is a derivative with respect to the $t$-coordinate and the auxiliary quantities are the following:
\ben
L_m(y) &=& \frac{(2r_++1+y)^{d-3}\left[d-2-\frac{(d-4)(2r_++1+y)}{1-y}\right]}{2(r_++1)(1-y)^{d-4}}\ f_1 f_2\ \p_y T_m(y) + \nonumber\\ 
&& + \frac{(2r_++1+y)^{d-2}}{2(r_++1)(1-y)^{d-4}} \Big[ (\p_yf_1 f_2 + f_1 \p_yf_2)\ \p_y T_m(y) + f_1 f_2\ \p^2_y T_m(y) \Big], \\
L_n(x) &=& \left(2-d\right) x\ \p_x T_n(x) + \left(1-x^2\right) \p_x^2 T_n(x).
\een
The equation \eqref{eqn:ev-coeff} was written at each of the collocation points \eqref{eqn:coll} separately. The resulting system of algebraic equations for the unknown $\varphi_{mn}$ was solved on each time step with the use of the LU decomposition method \cite{nrF}.

The pseudospectral method provides an exact solution at the collocation points. Since the overall solution is approximated by a set of polynomials, bumps can appear in the regions between the points. They are an artifact of the numerical scheme. The density of the grid, i.e., the amount of collocation points used in our setup ($50$ in each spatial direction), ensured a very good accuracy of the obtained outcomes and facilitated their interpretation. It was also sufficient for modeling the step-type profile of the scalar field~\eqref{boo}, which was necessary for reflecting the existence of the brane in the spacetime.


\subsection{Dynamical behavior}
\label{ssec:dyn-results}

The location of the brane in the spacetime was determined taking into account the values of the energy density of the scalar field $\Phi$
\be
T_{0}^{\; 0} = \frac{1}{f_1 f_2^{\; 1 - \ga (d-3)}} \big( \p_{t} \Phi \big)^2
+ \frac{1}{4} f_1 f_2^{\; 1 - \ga} \big( \p_{r} \Phi \big)^2 + \frac{1}{r^2} \big( \p_{\theta} \Phi \big)^2
+\frac{1}{2} V(\Phi).
\label{eqn:en-dens}
\ee
The results of the computations are presented through series of snapshots of the temporal evolution of the brane-black hole system. Each of the plots shows the energy density of the scalar field \eqref{eqn:en-dens} within the black hole spacetime in cylindrical coordinates. The locations of the black hole horizons and the number of the evolution time step are shown on each diagram.

The investigated spacetime dimensions were $4$, $5$, $6$, $7$ and $8$. The brane thickness $l$ was set as $0.1$, $0.3$ and $0.4$, while its initial velocity $v$ was assumed to be equal to $0.1$, $0.3$ and $0.7$. 
Similarly to the case of static configurations of the BDBH system described in section~\ref{sec:sta}, three values of the dilatonic constant $\alpha$ were considered, namely $-\sqrt{3}$, $-1$ and $0$. The constant $\eta$ was equal to $0.1$ in all simulations, which in addition to the value of $l$ determined the tension of the domain wall in each of the investigated cases.

\begin{figure}[h]
\centering
(a)\\
\includegraphics[scale=0.6]{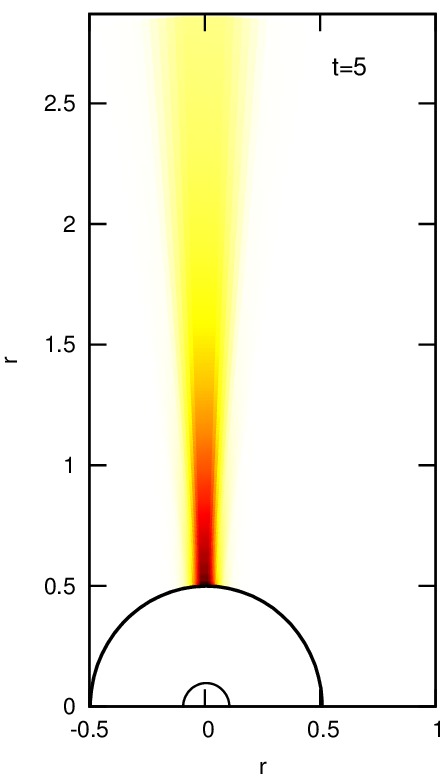}
\includegraphics[scale=0.6]{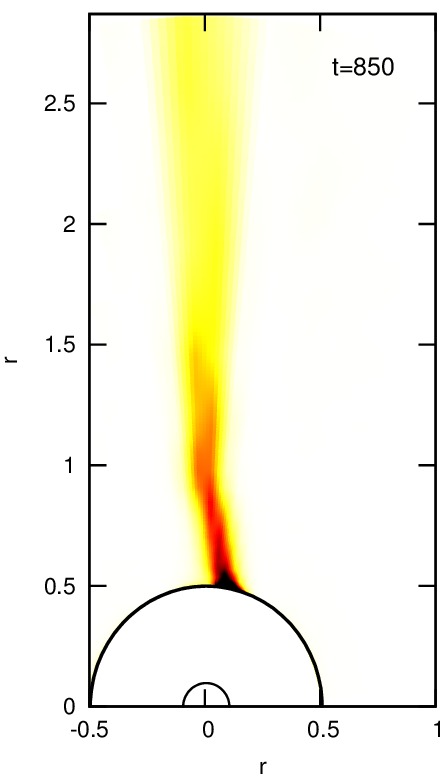}
\includegraphics[scale=0.6]{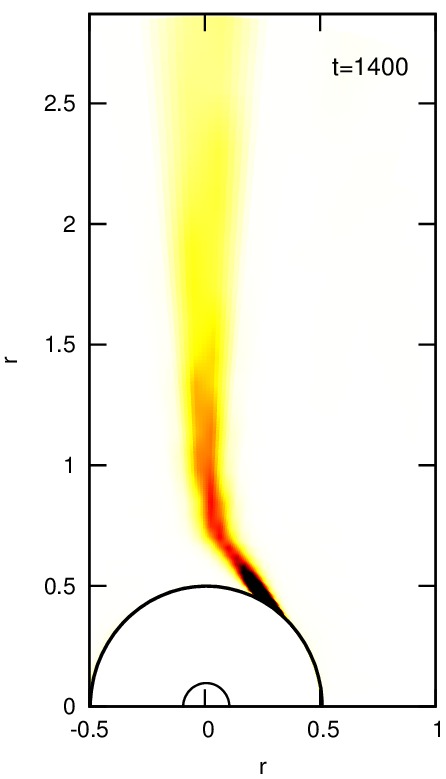}
\includegraphics[scale=0.6]{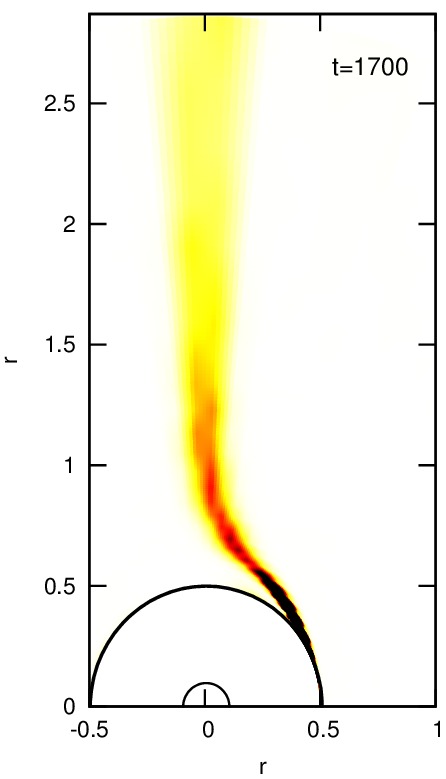}
\includegraphics[scale=0.6]{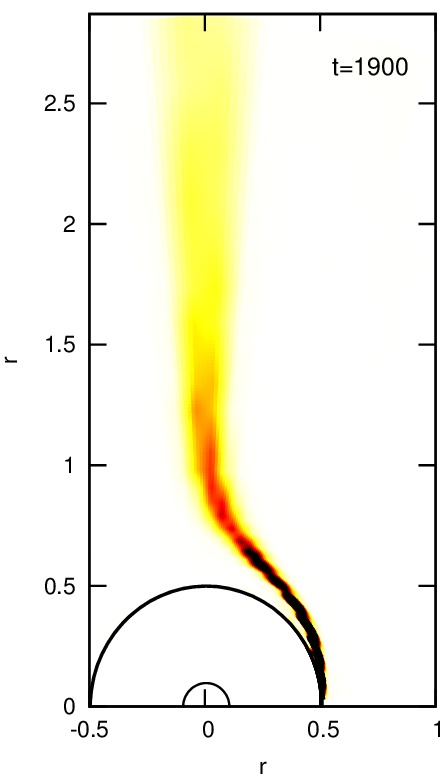}
\includegraphics[scale=0.6]{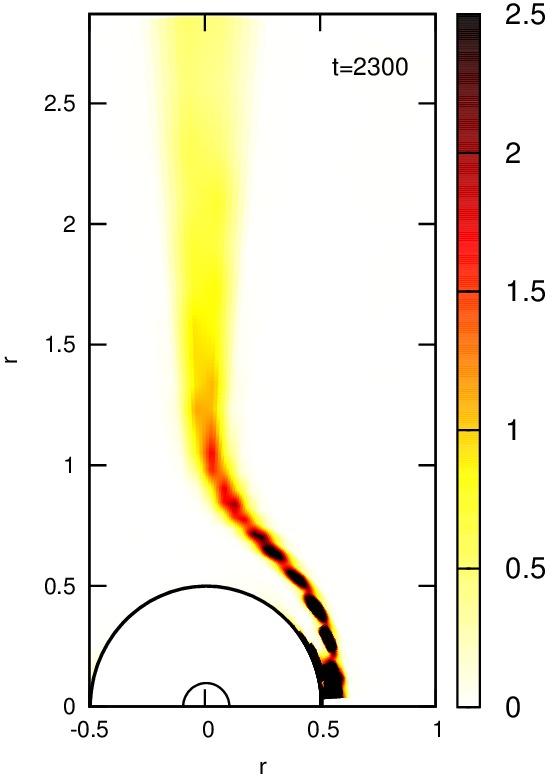}\\
(b)\\
\includegraphics[scale=0.6]{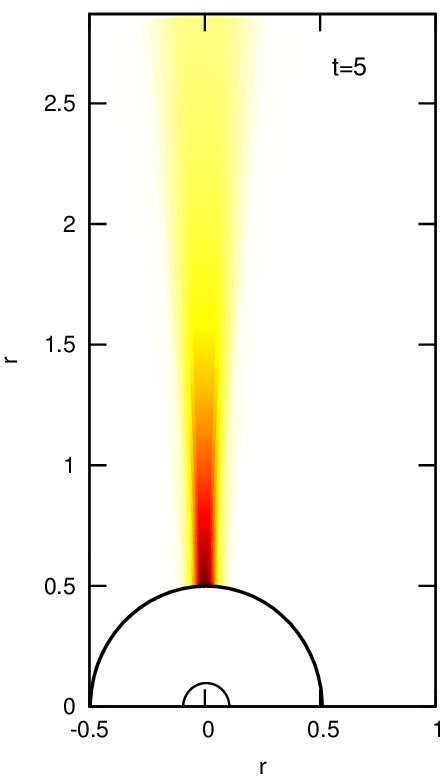}
\includegraphics[scale=0.6]{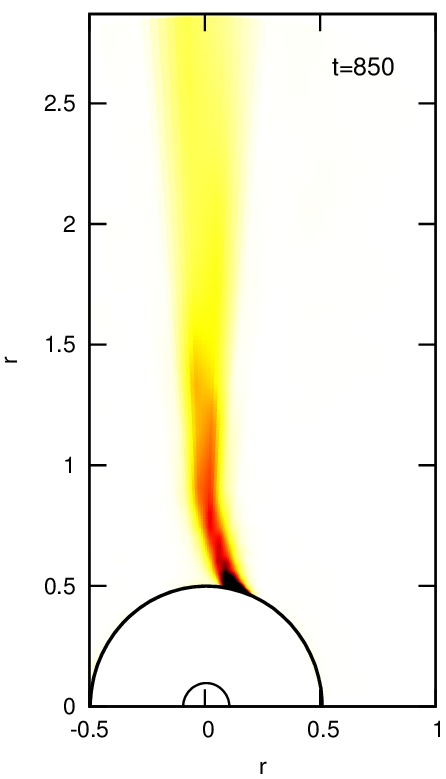}
\includegraphics[scale=0.6]{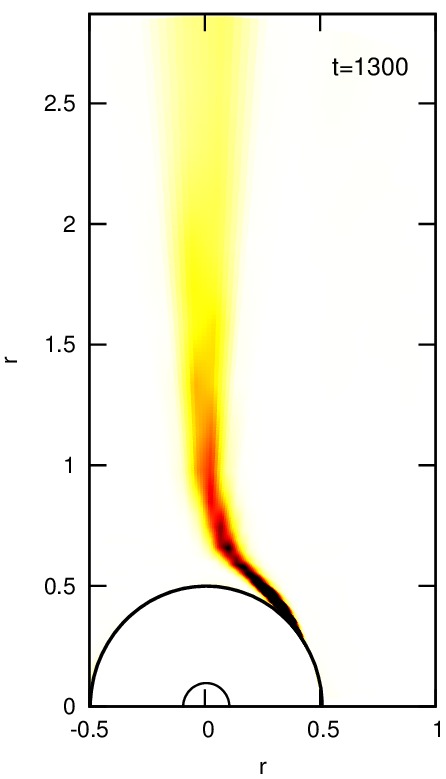}
\includegraphics[scale=0.6]{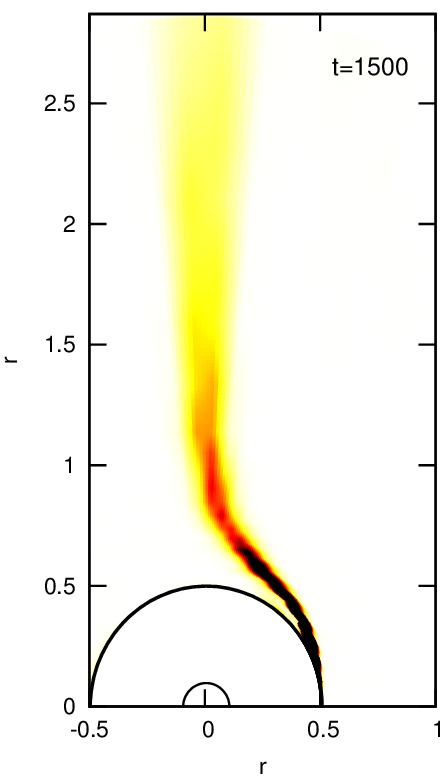}
\includegraphics[scale=0.6]{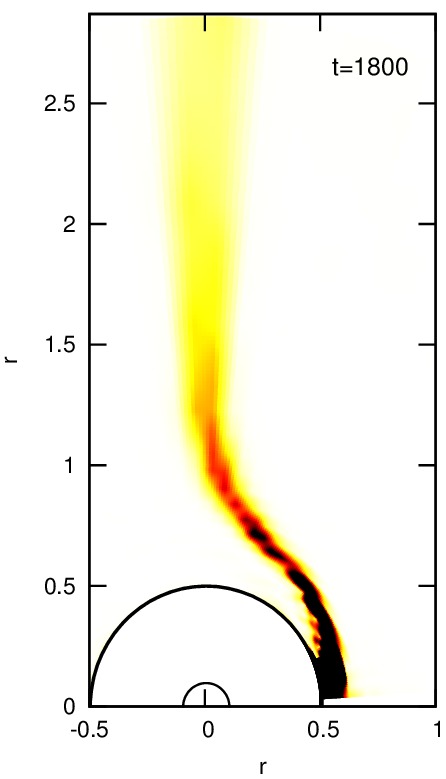}
\includegraphics[scale=0.6]{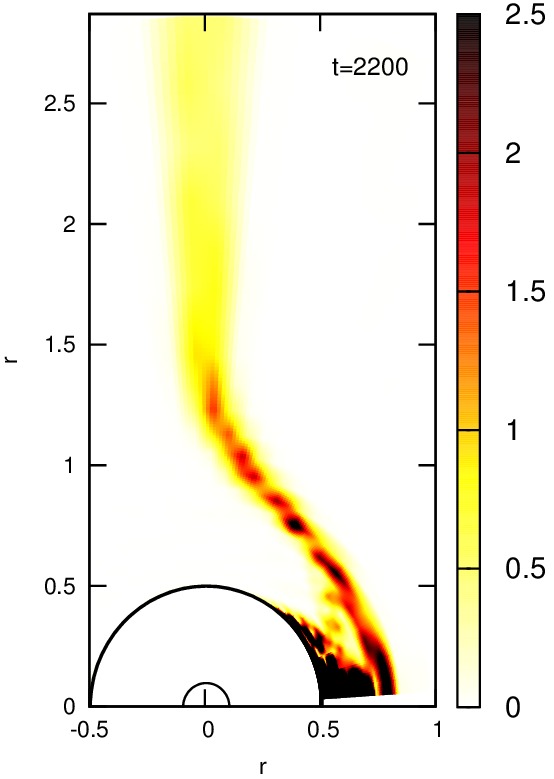}\\
(c)\\
\includegraphics[scale=0.6]{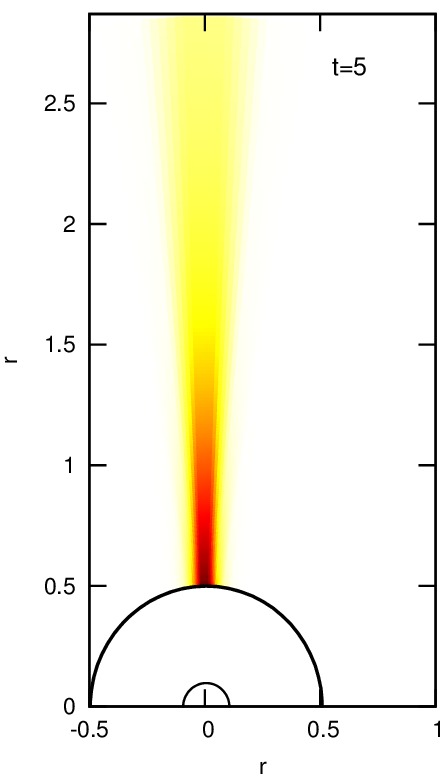}
\includegraphics[scale=0.6]{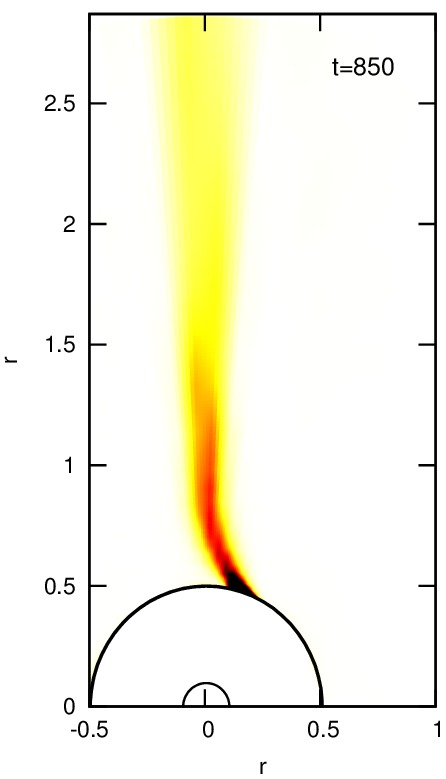}
\includegraphics[scale=0.6]{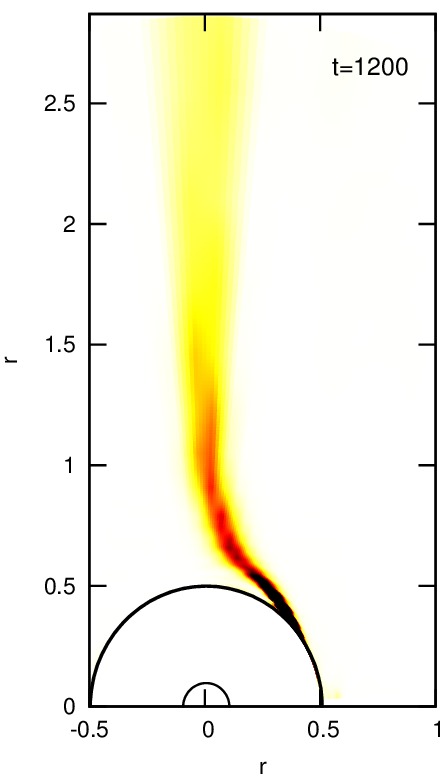}
\includegraphics[scale=0.6]{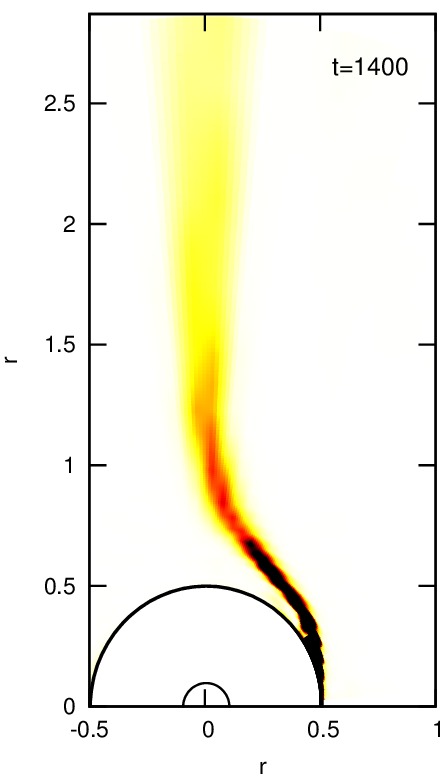}
\includegraphics[scale=0.6]{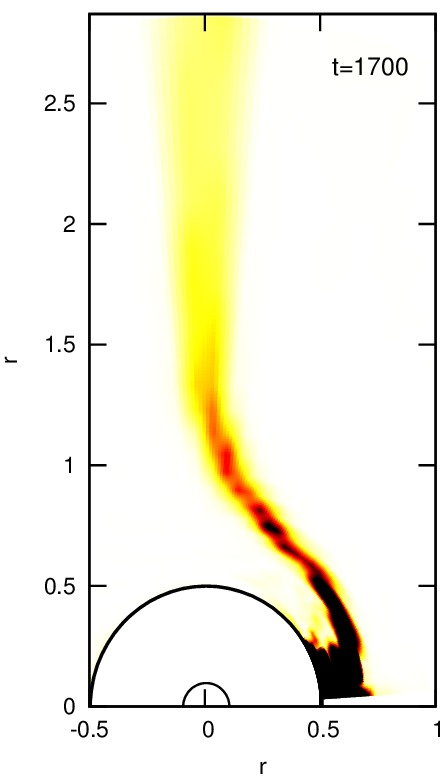}
\includegraphics[scale=0.6]{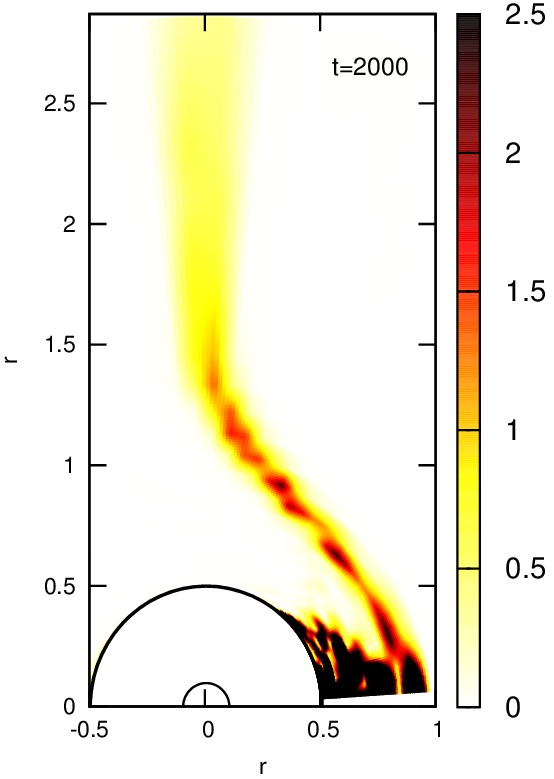}\\
(d)\\
\includegraphics[scale=0.6]{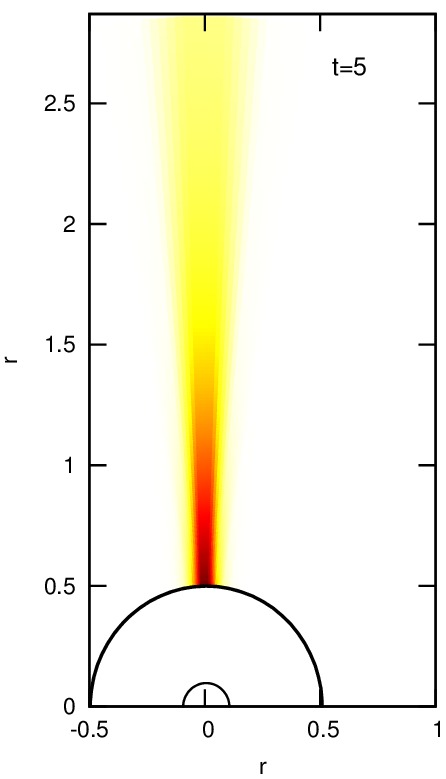}
\includegraphics[scale=0.6]{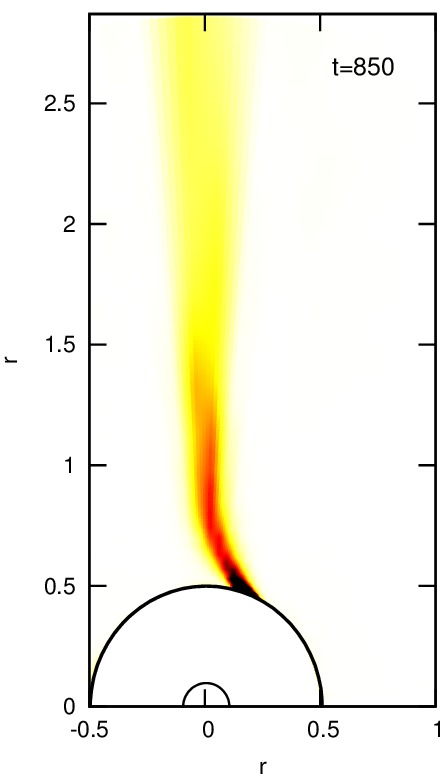}
\includegraphics[scale=0.6]{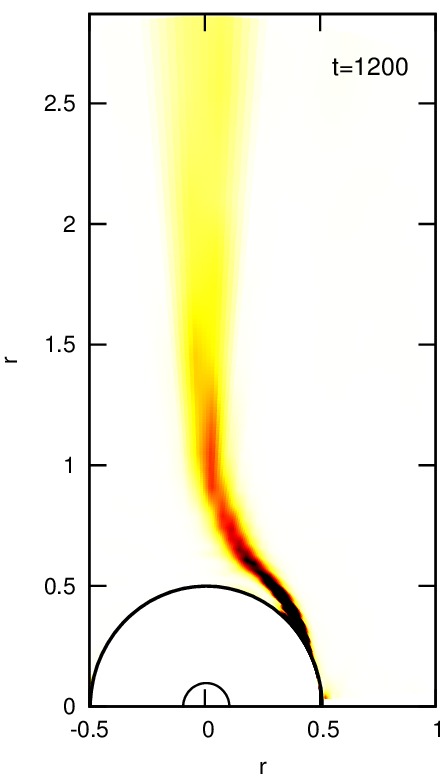}
\includegraphics[scale=0.6]{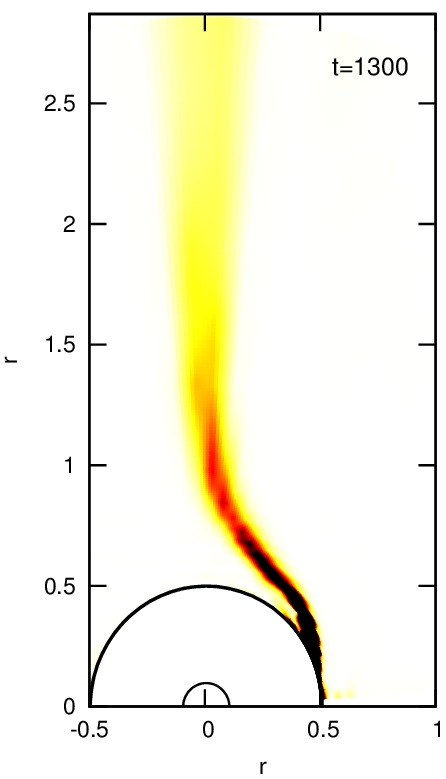}
\includegraphics[scale=0.6]{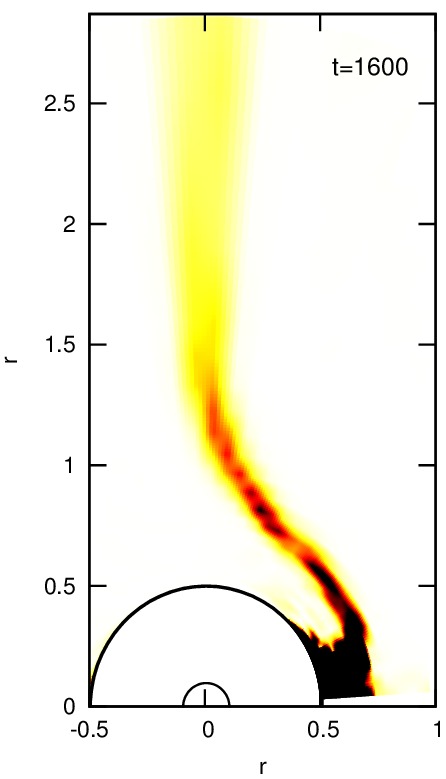}
\includegraphics[scale=0.6]{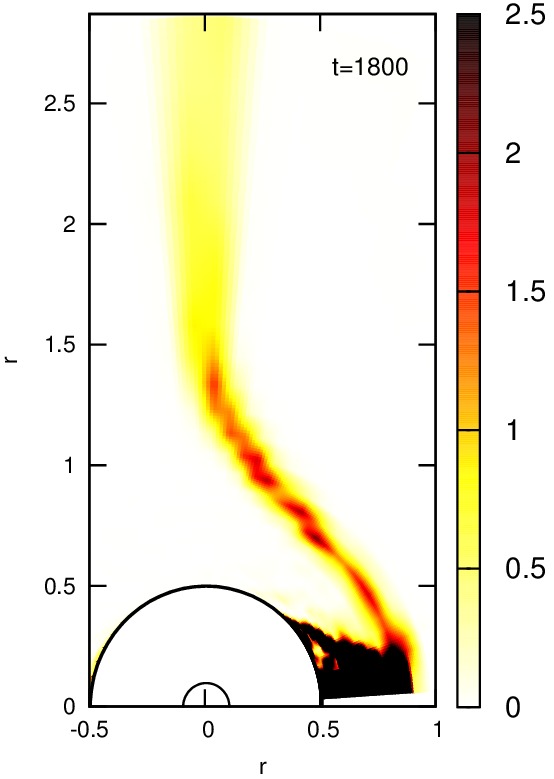}
\caption{The evolution of the BDBH system with the nonextremal black hole for $d$ equal to (a)~$4$, (b)~$5$, (c)~$6$ and (d)~$7$. The remaining parameters were set as $l=0.1$, $v=0.1$ and $\alpha=-1$.}
\label{fig:dyn-nonextr-d}
\end{figure}

\begin{figure}[h]
\centering
(a)\\
\includegraphics[scale=0.6]{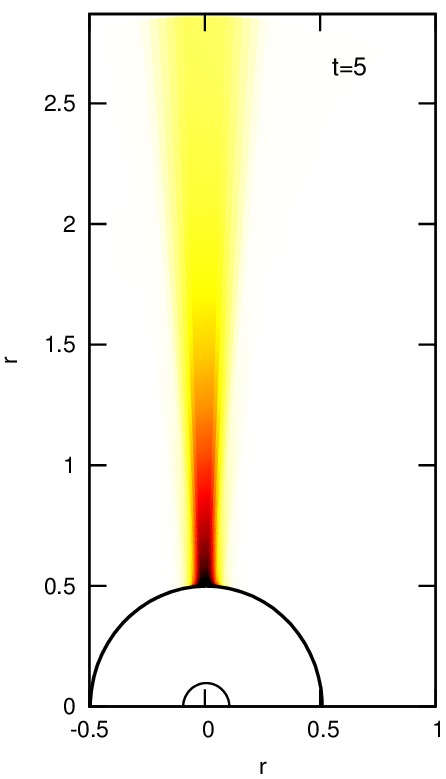}
\includegraphics[scale=0.6]{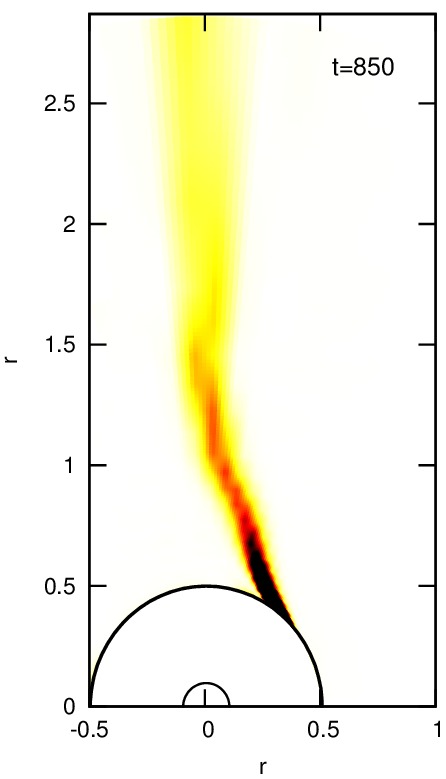}
\includegraphics[scale=0.6]{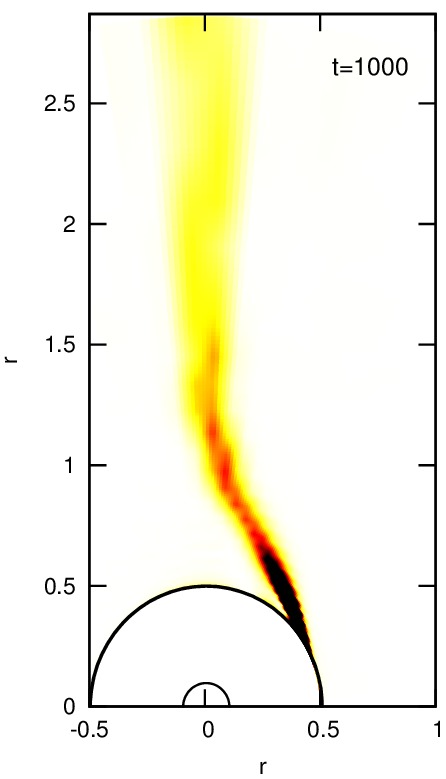}
\includegraphics[scale=0.6]{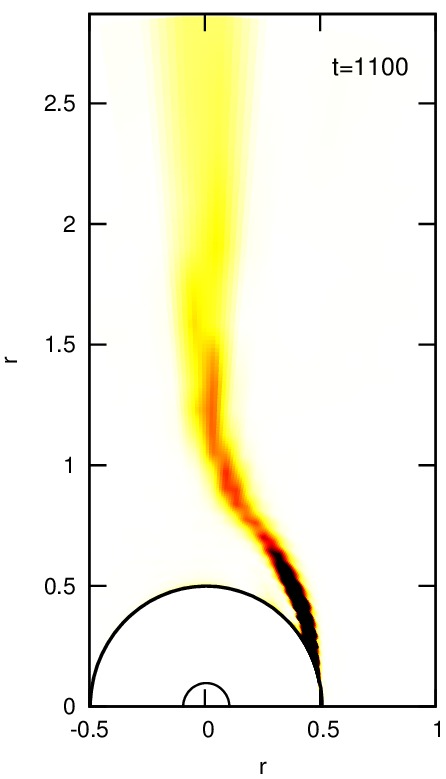}
\includegraphics[scale=0.6]{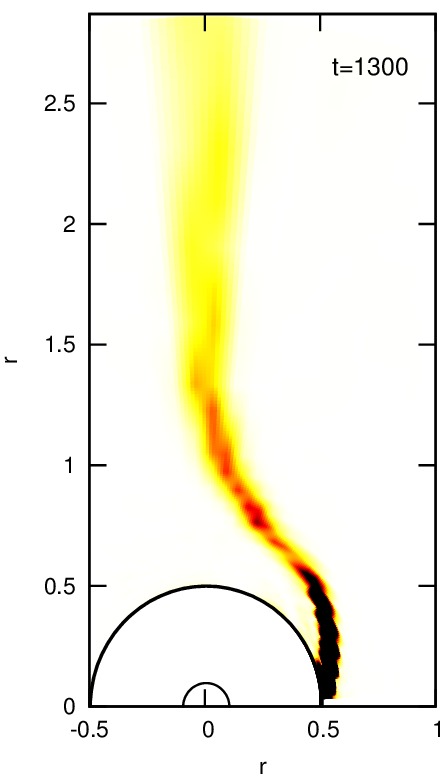}
\includegraphics[scale=0.6]{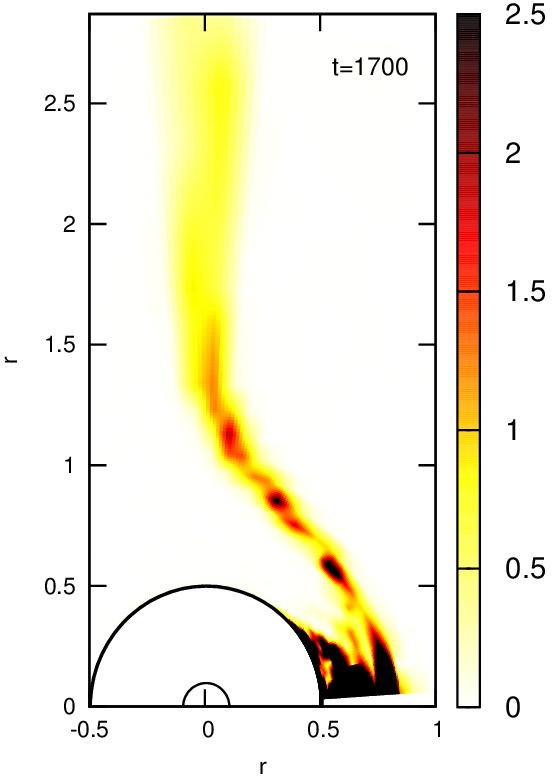}\\
(b)\\
\includegraphics[scale=0.6]{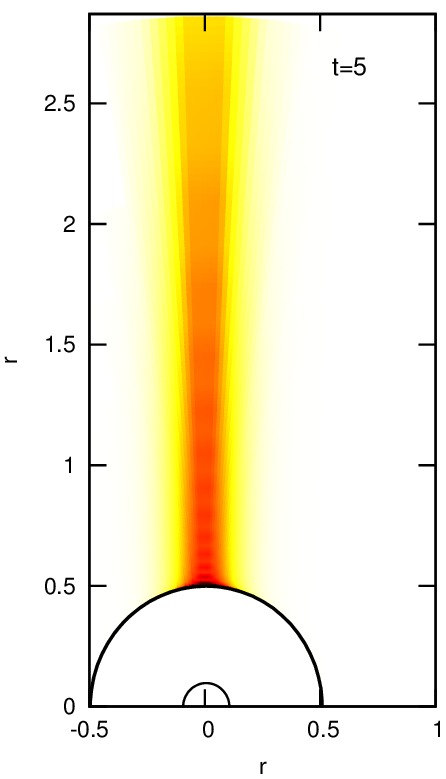}
\includegraphics[scale=0.6]{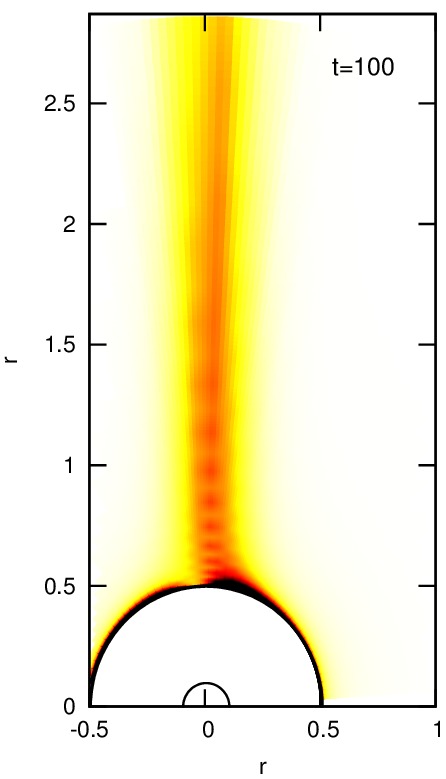}
\includegraphics[scale=0.6]{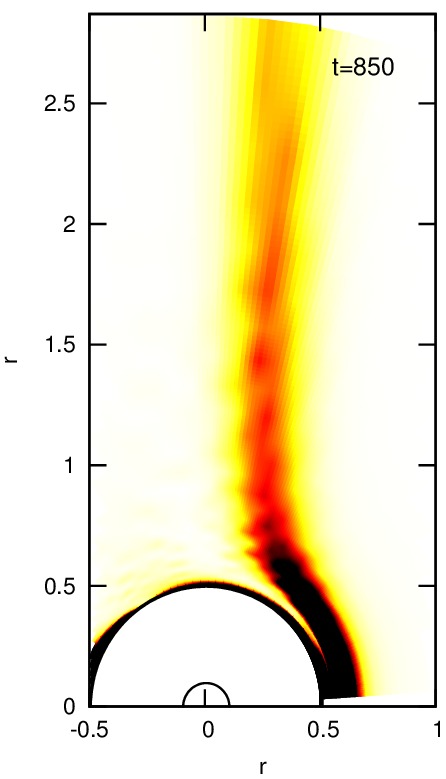}
\includegraphics[scale=0.6]{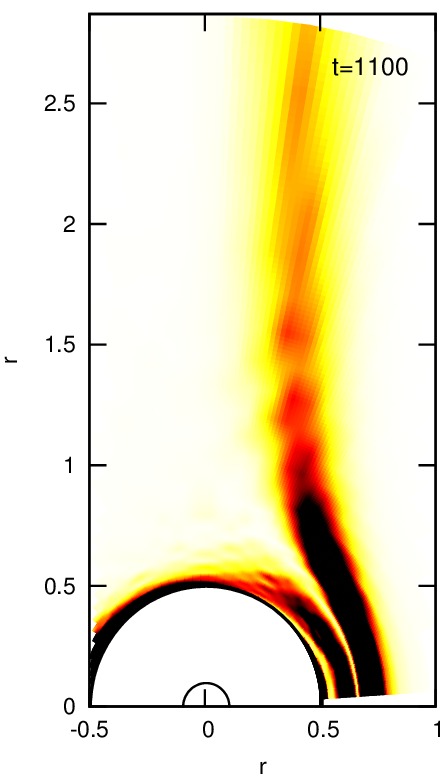}
\includegraphics[scale=0.6]{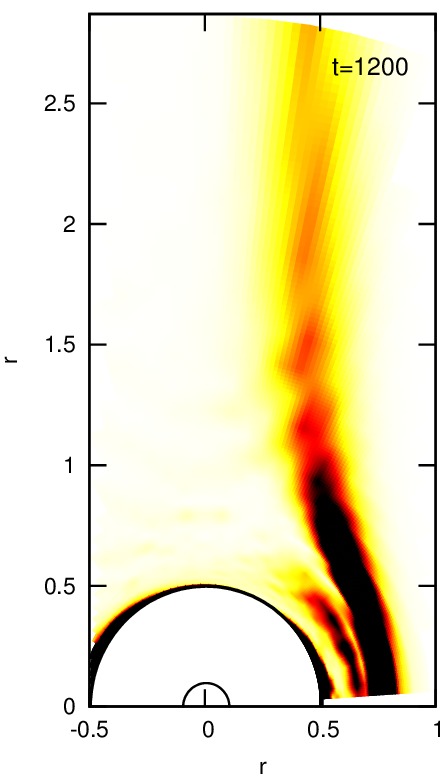}
\includegraphics[scale=0.6]{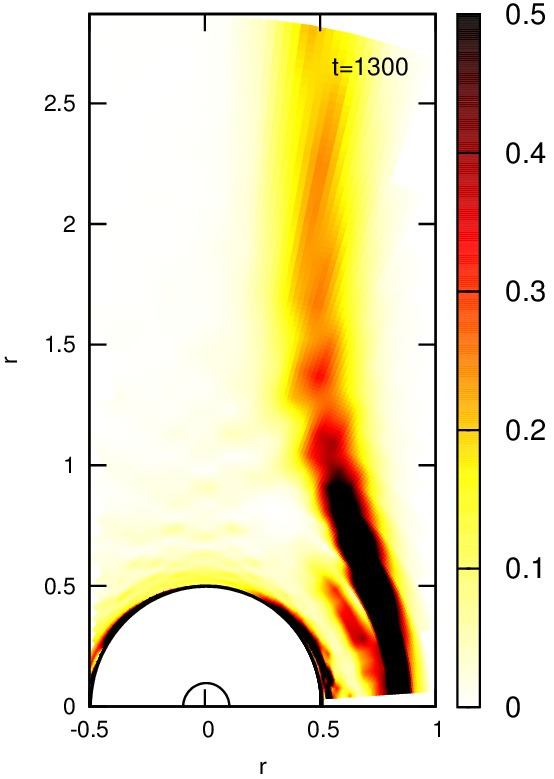}\\
(c)\\
\includegraphics[scale=0.6]{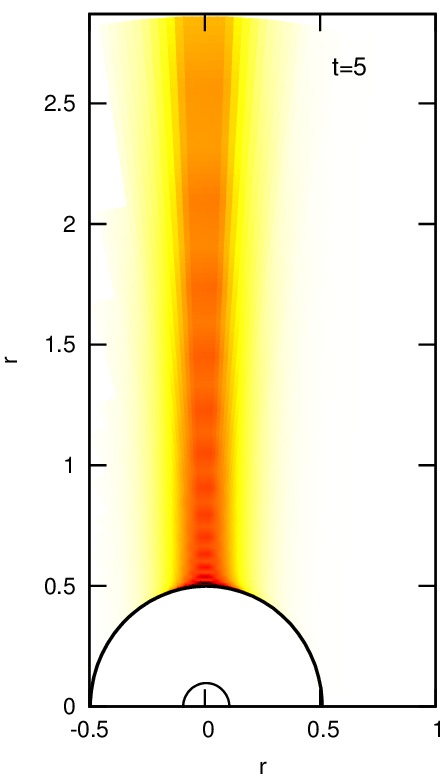}
\includegraphics[scale=0.6]{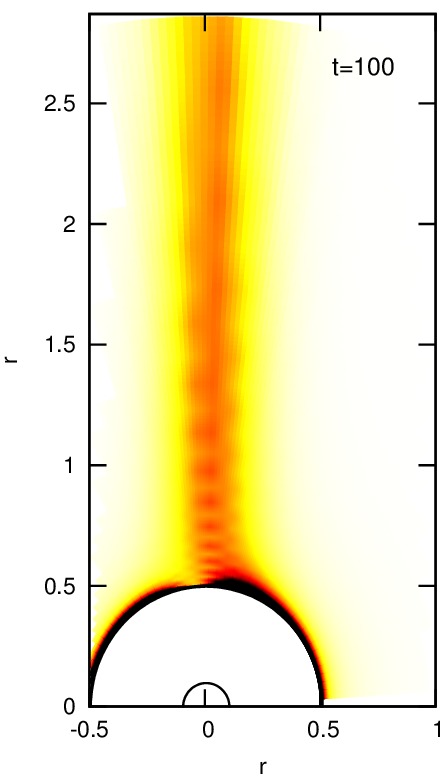}
\includegraphics[scale=0.6]{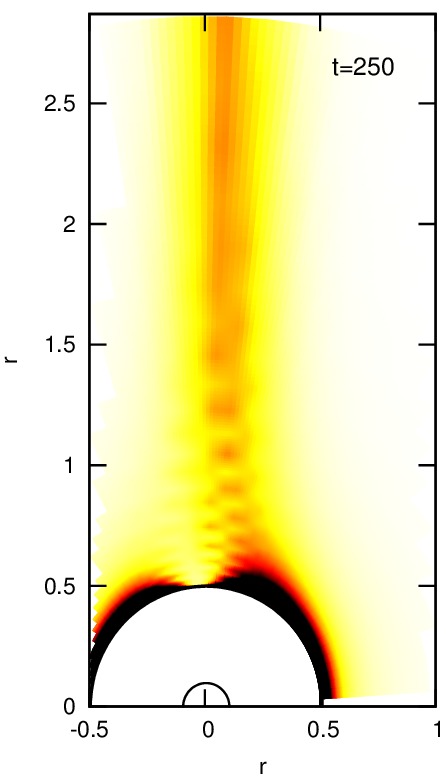}
\includegraphics[scale=0.6]{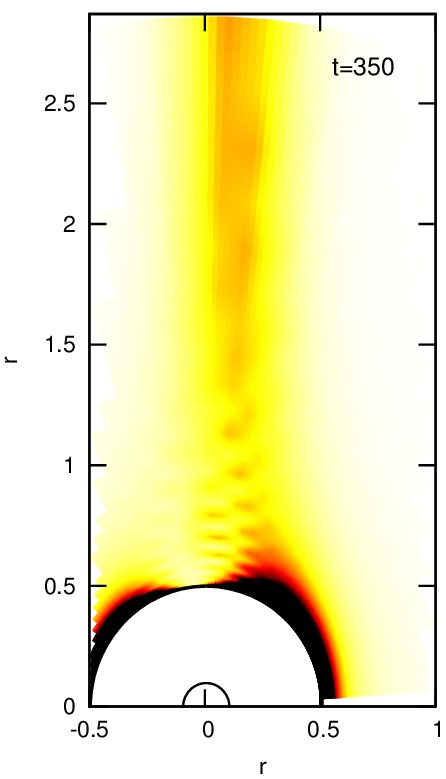}
\includegraphics[scale=0.6]{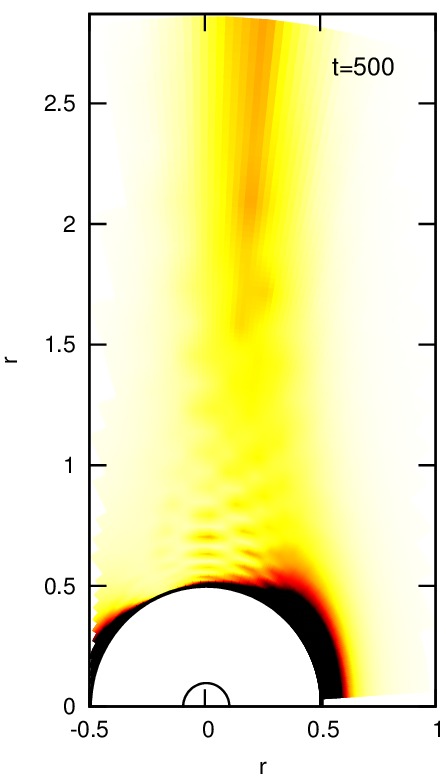}
\includegraphics[scale=0.6]{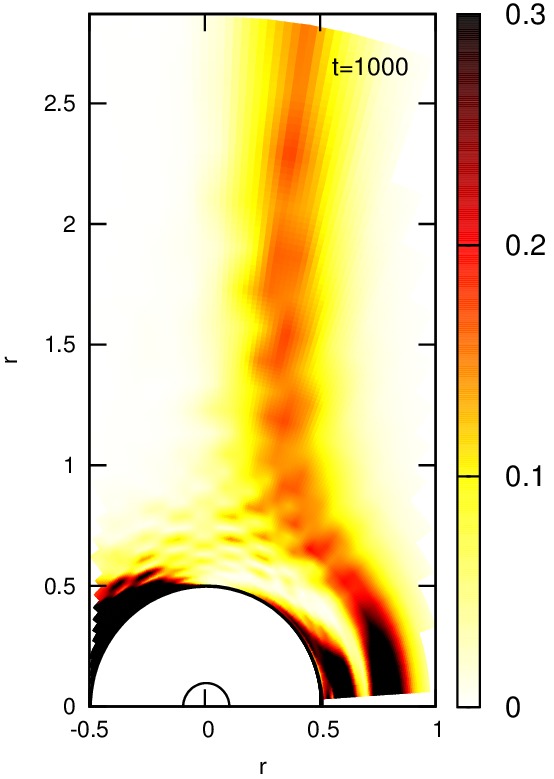}
\caption{The evolution of the BDBH system with the nonextremal black hole for $l$ equal to (a)~$0.1$, (b)~$0.3$ and (c)~$0.4$. The remaining parameters were set as $d=5$, $v=0.3$ and $\alpha=0$.}
\label{fig:dyn-nonextr-l}
\end{figure}

\begin{figure}[h]
\centering
(a)\\
\includegraphics[scale=0.6]{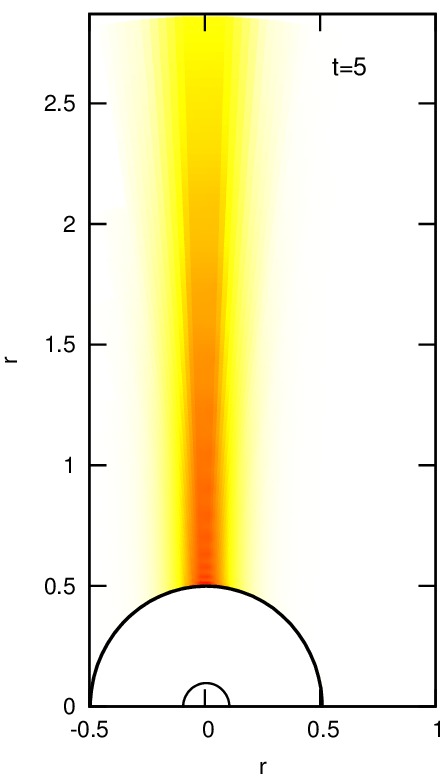}
\includegraphics[scale=0.6]{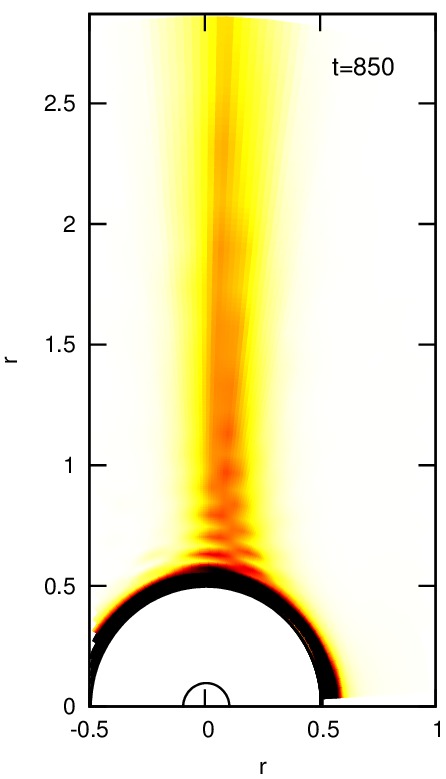}
\includegraphics[scale=0.6]{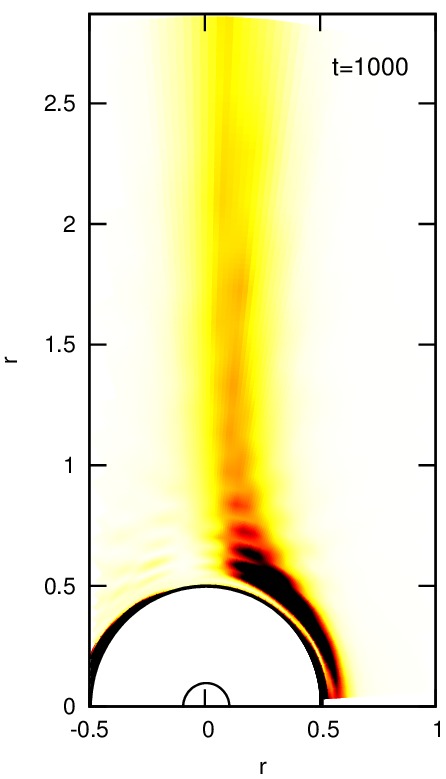}
\includegraphics[scale=0.6]{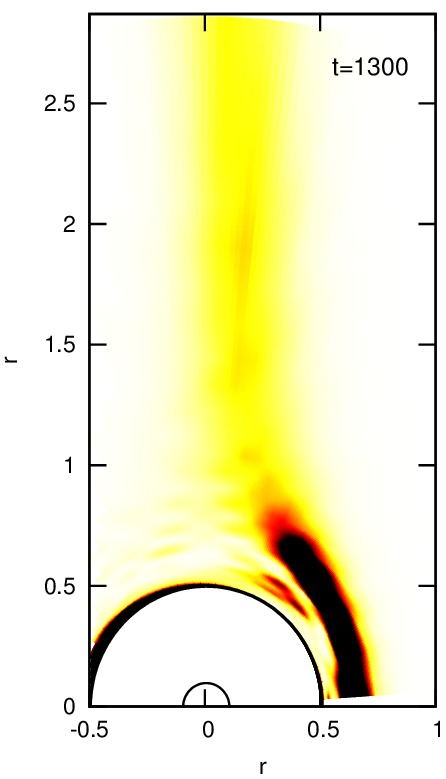}
\includegraphics[scale=0.6]{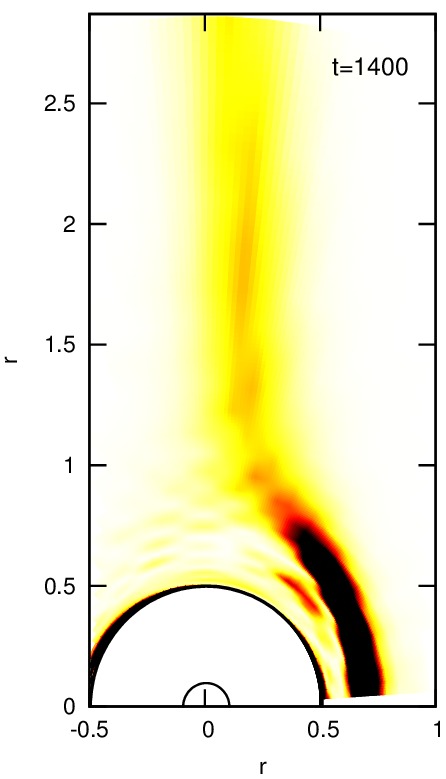}
\includegraphics[scale=0.6]{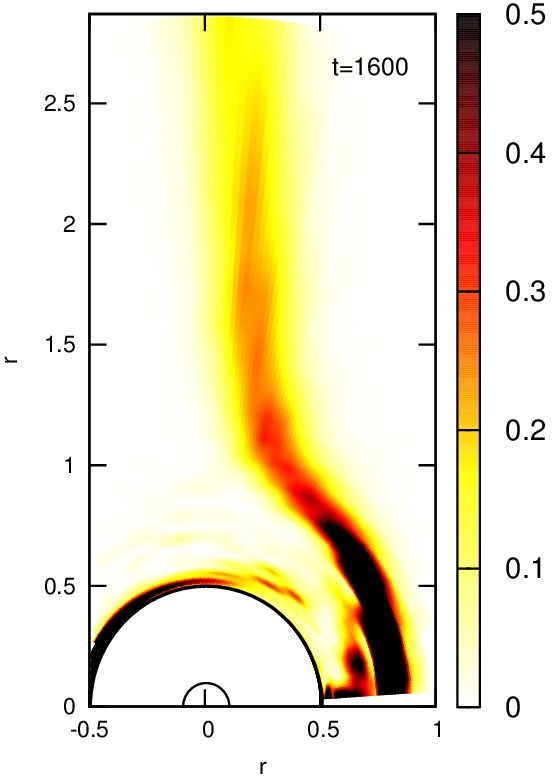}\\
(b)\\
\includegraphics[scale=0.6]{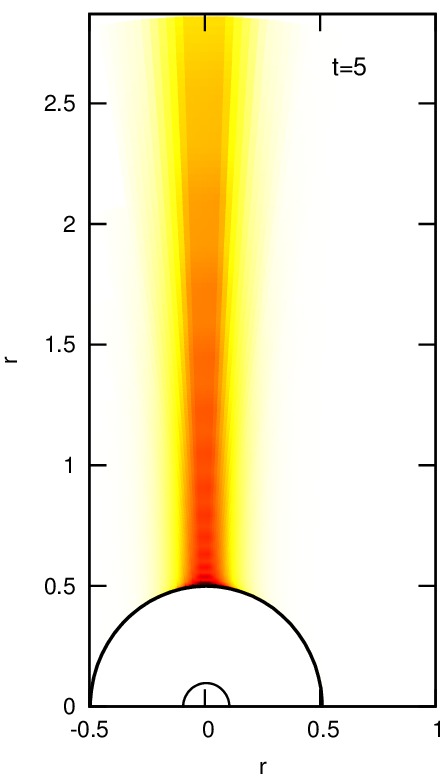}
\includegraphics[scale=0.6]{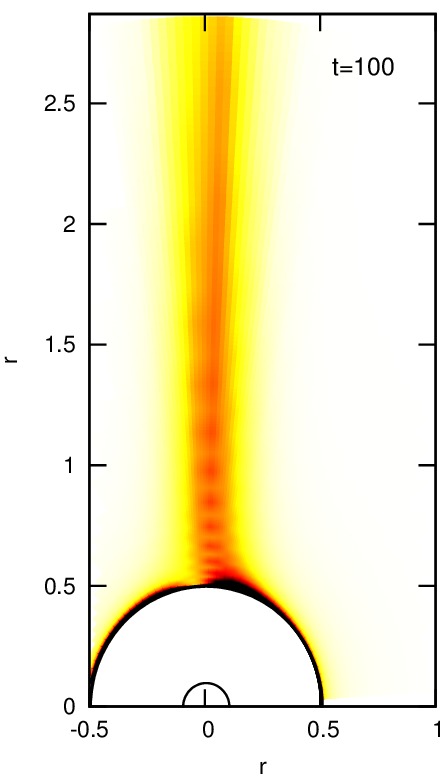}
\includegraphics[scale=0.6]{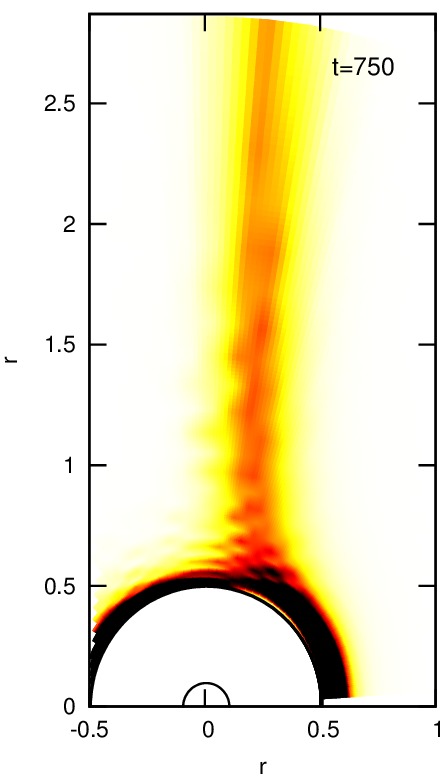}
\includegraphics[scale=0.6]{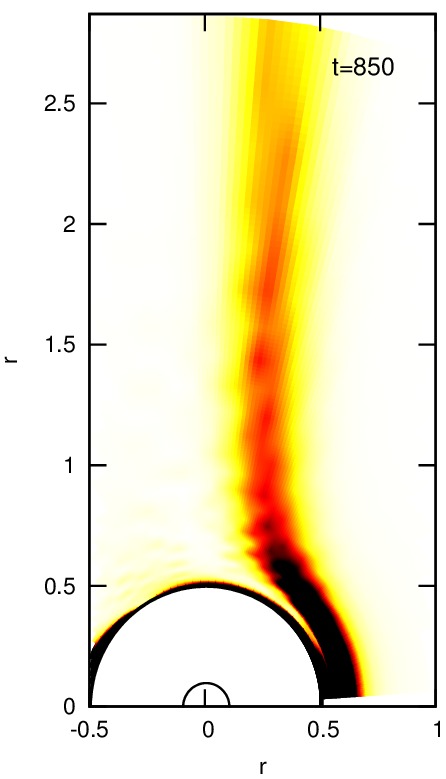}
\includegraphics[scale=0.6]{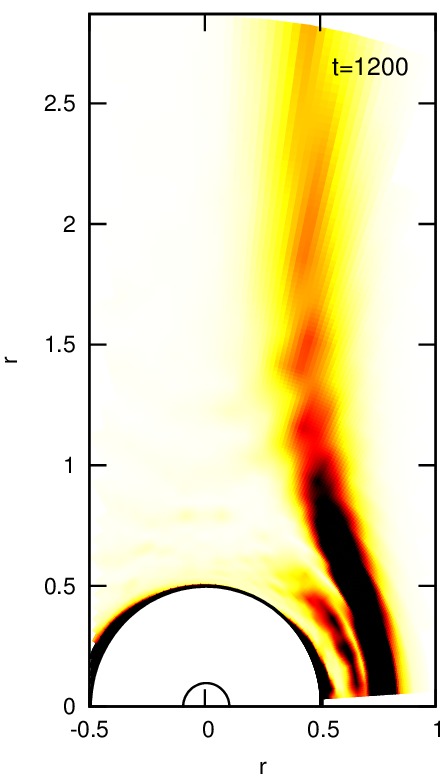}
\includegraphics[scale=0.6]{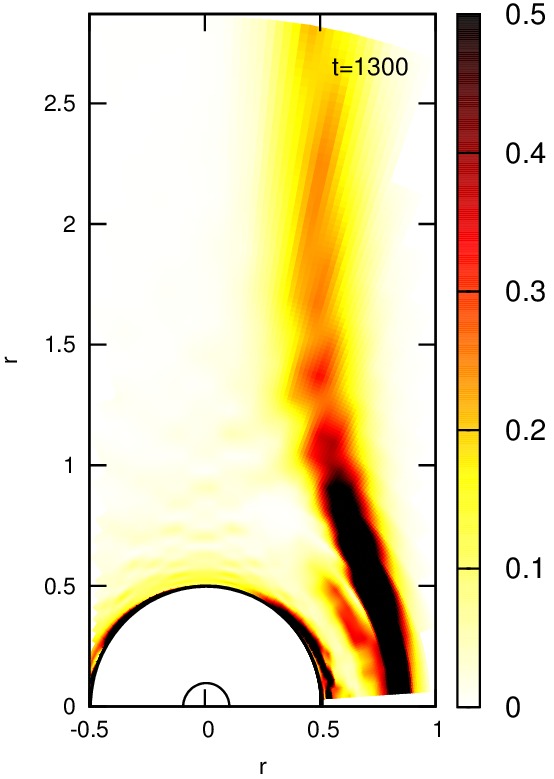}\\
(c)\\
\includegraphics[scale=0.6]{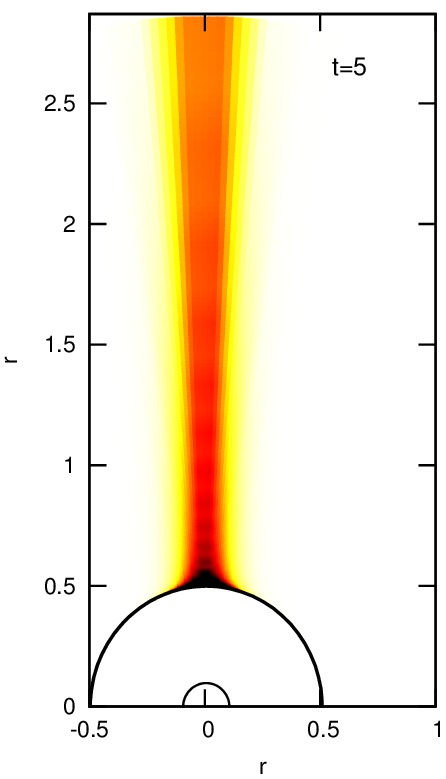}
\includegraphics[scale=0.6]{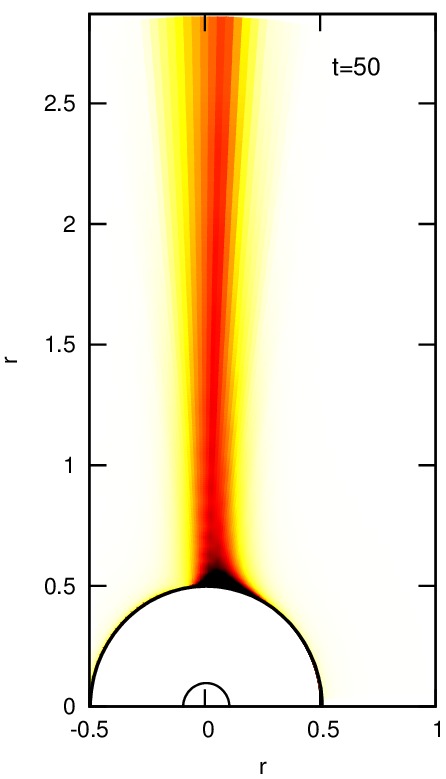}
\includegraphics[scale=0.6]{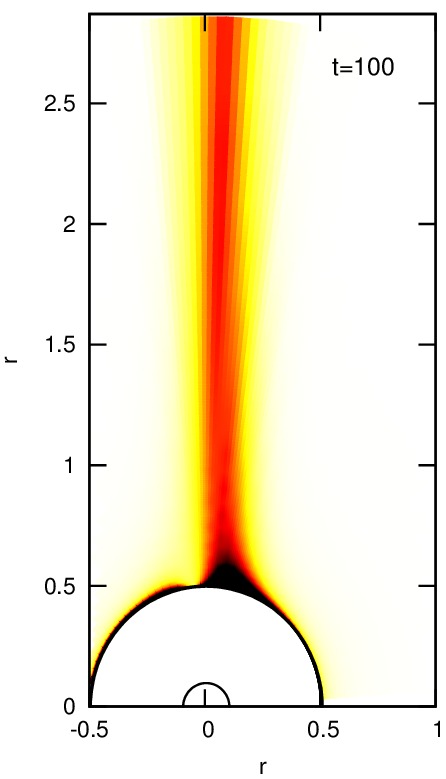}
\includegraphics[scale=0.6]{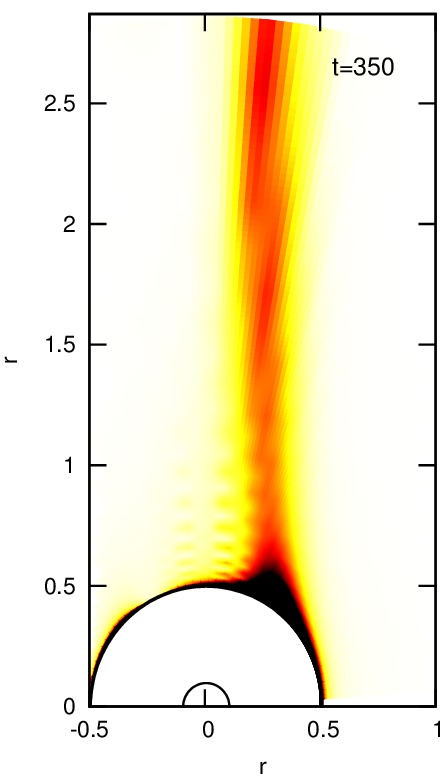}
\includegraphics[scale=0.6]{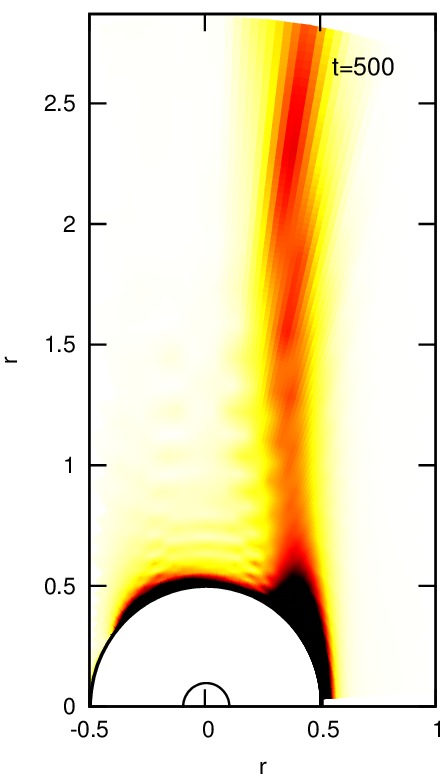}
\includegraphics[scale=0.6]{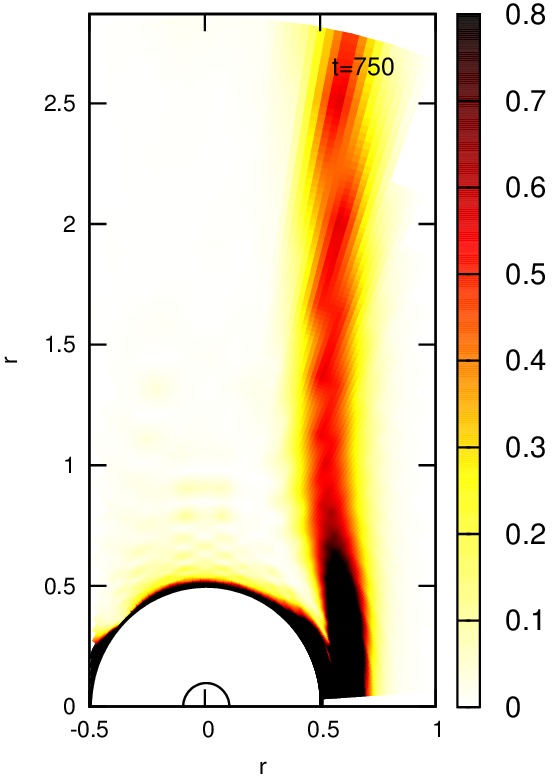}
\caption{The evolution od the BDBH system with the nonextremal black hole for $v$ equal to (a)~$0.1$, (b)~$0.3$ and (c)~$0.7$. The remaining parameters were set as $d=5$, $l=0.3$ and $\alpha=-1$.}
\label{fig:dyn-nonextr-v}
\end{figure}

\begin{figure}[h]
\centering
(a)\\
\includegraphics[scale=0.6]{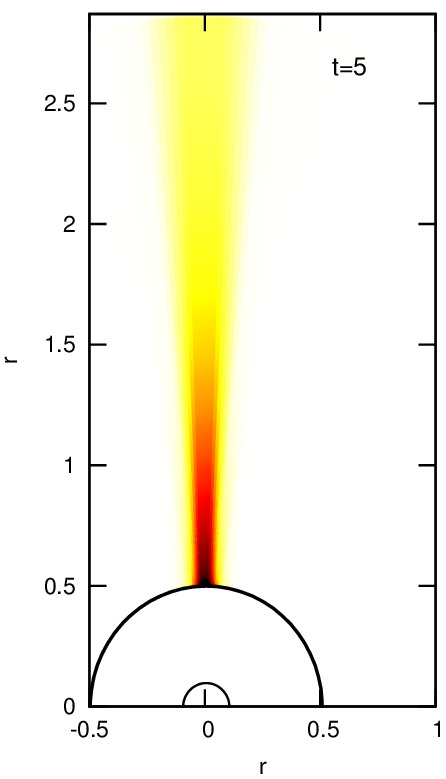}
\includegraphics[scale=0.6]{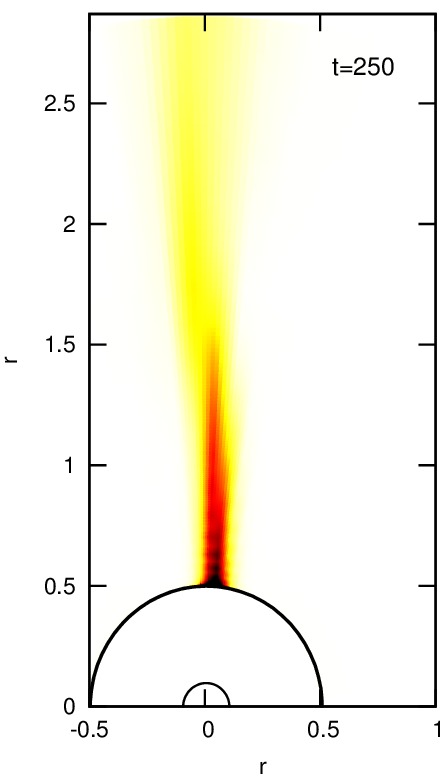}
\includegraphics[scale=0.6]{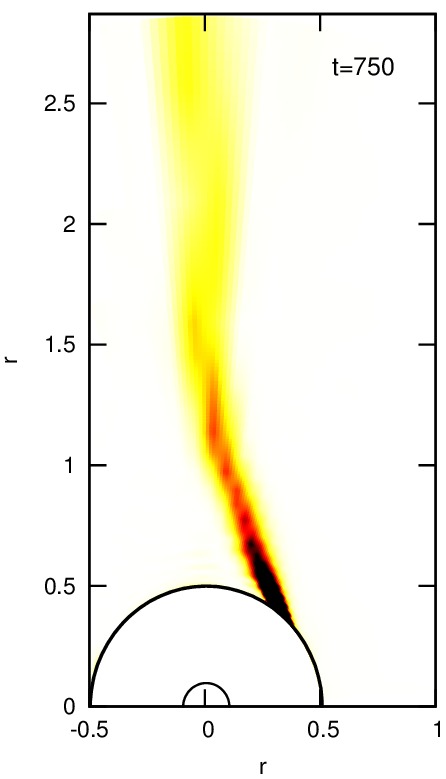}
\includegraphics[scale=0.6]{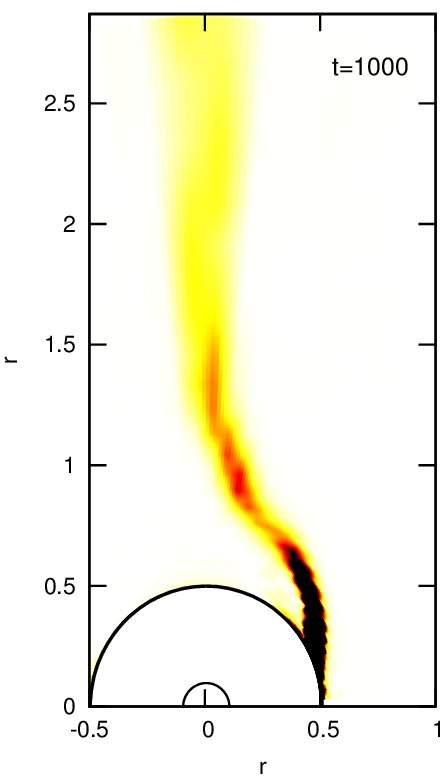}
\includegraphics[scale=0.6]{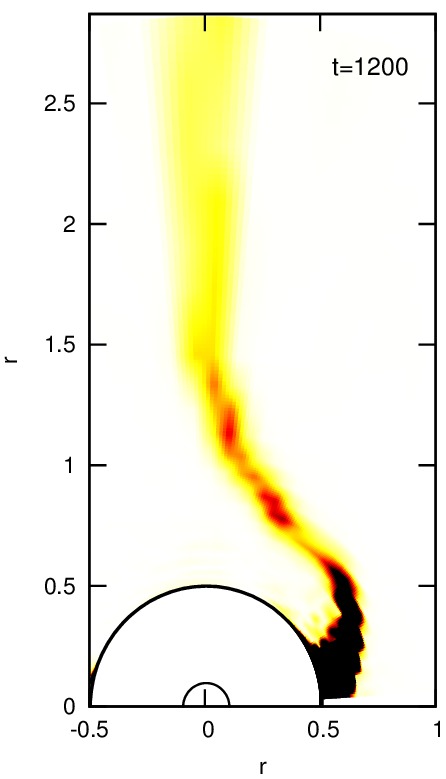}
\includegraphics[scale=0.6]{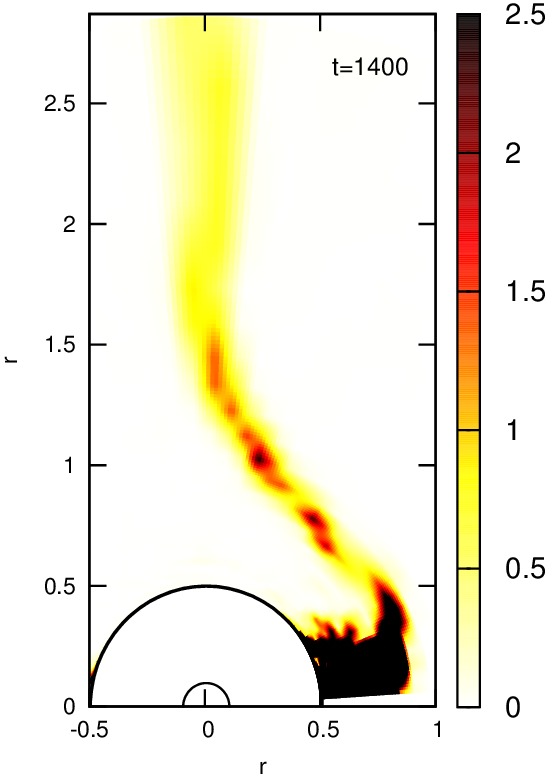}\\
(b)\\
\includegraphics[scale=0.6]{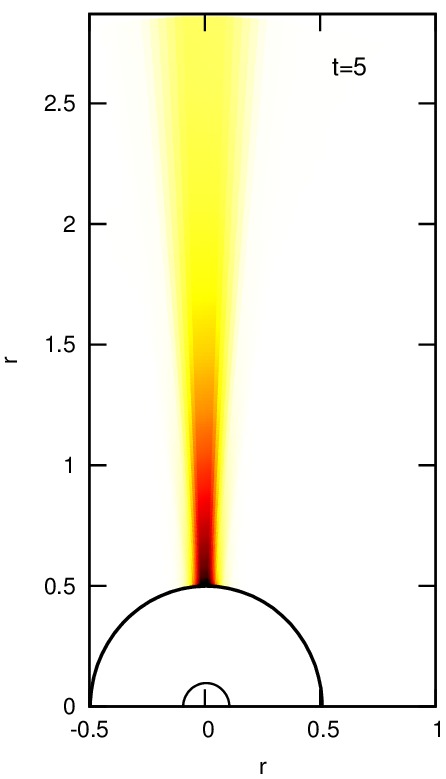}
\includegraphics[scale=0.6]{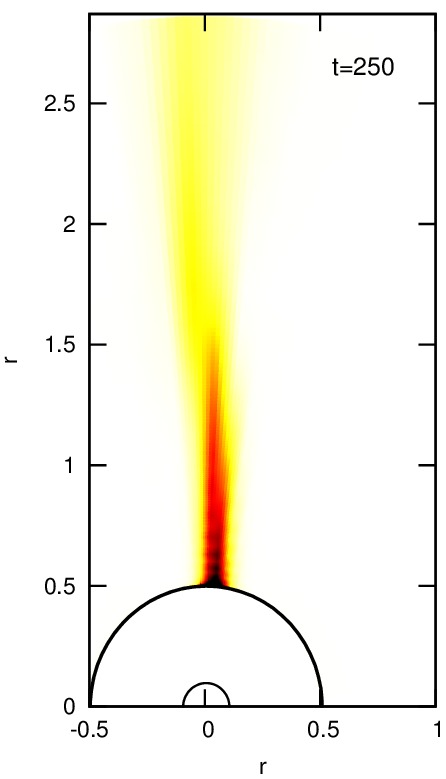}
\includegraphics[scale=0.6]{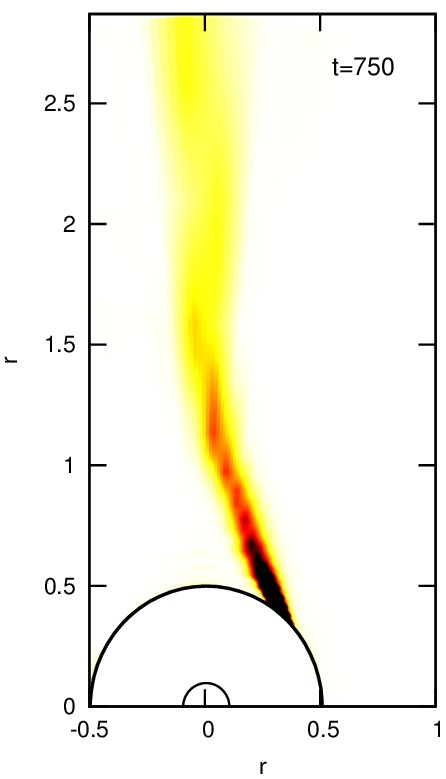}
\includegraphics[scale=0.6]{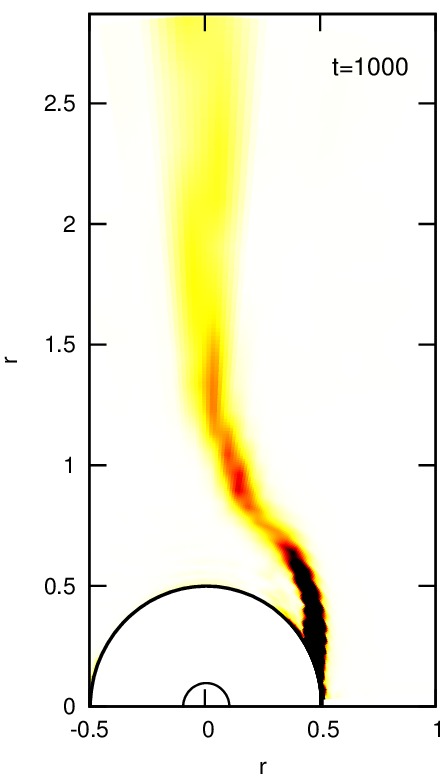}
\includegraphics[scale=0.6]{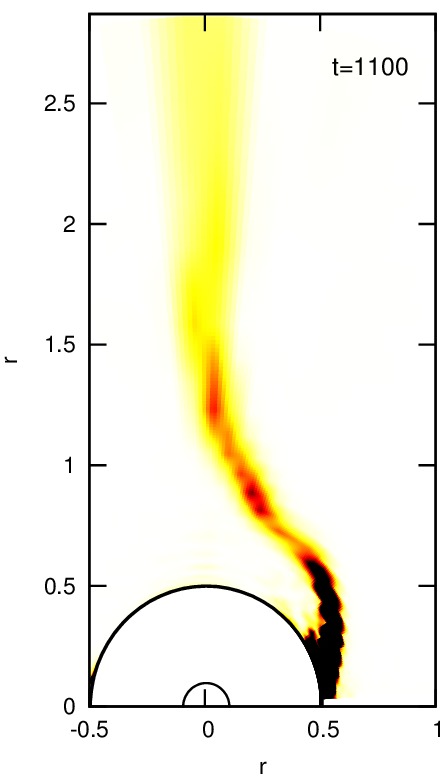}
\includegraphics[scale=0.6]{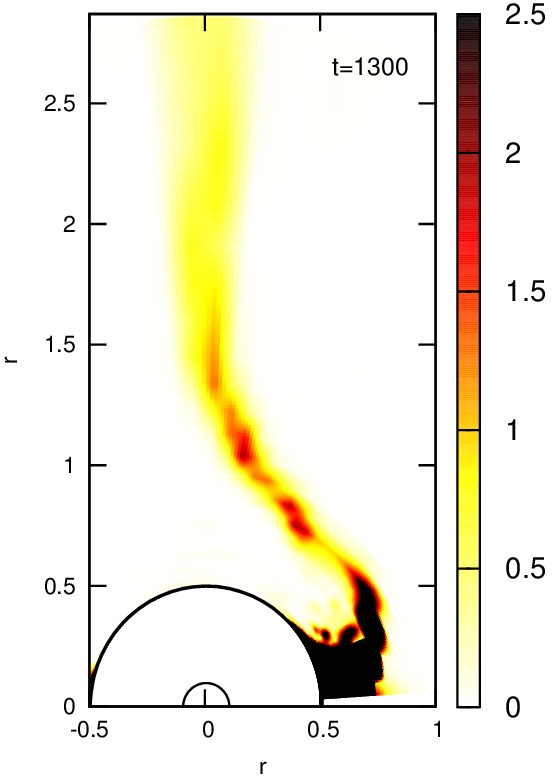}
\caption{The evolution of the BDBH system with the nonextremal black hole for $\alpha$ equal to (a)~$-1$ and (b)~$0$. The remaining parameters were set as $d=7$, $l=0.1$ and $v=0.3$.}
\label{fig:dyn-nonextr-alfa}
\end{figure}

Figures~\ref{fig:dyn-nonextr-d}--\ref{fig:dyn-nonextr-alfa} present the dependence of the dynamical behavior of the brane-black hole system on $d$, $l$, $v$ and $\alpha$, respectively, in the case of a nonextremal black hole with $r_-=0.1$ and $r_+=0.5$. In all cases, the plane within which the brane was located initially divided the black hole into two hemispheres. Since the brane acquired a non-zero velocity at the beginning of the investigated process, with the passage of time it moved in a specific direction. Such a situation can also be interpreted as a black hole being expelled from the brane.

The research conducted for a BDBH system containing a nonextremal black hole revealed that the expulsion appears earlier for larger $d$ and the black hole is expelled earlier from a thicker brane (with bigger $l$). The radius of the area of the brane deformation is bigger for a thicker brane, it moves away from the initial position within a larger distance from the black hole. The energy density associated with a thick brane is smaller in comparison to a thin brane. Bigger initial velocity of the brane results in an earlier expulsion of the black hole and causes an increase of the radius of the area of the brane deformation nearby the black hole. The dynamics of the studied brane-black hole system does not depend on the value of $\alpha$.

The evolution of the examined system was also followed for an extremal black hole with $r_-=r_+=0.5$. The outcomes were qualitatively the same as in the nonextremal case, the only 
difference was that the expulsion appeared a bit earlier in the extremal case.

\section{Conclusions}
\label{sec:conclusions}

In our paper we pay attention to static configurations and a dynamical evolution of the system consisting of a higher-dimensional spherically symmetric dilaton black hole and a Dirac-Goto-Nambu type brane.
Due to the employed theoretical descriptions of the brane in the static and dynamical cases, the 
former analysis corresponds to an infinitely thin brane, while the latter covers also branes of 
non-negligible thicknesses.

It has been established \cite{gar91,cam90,sha91,hol92}
that the classical black hole solutions of Einstein-Maxwell gravity
have quite new features when the theory is modified by the introduction of the low-energy string corrections
to the underlying action. The key property of the aforementioned actions is bounded with the nontrivial
coupling of the dilaton field with gravity and the field strength of Maxwell or Yang-Mills gauge fields.
The charged black hole or black brane solutions in the low-energy string theory differ to the great
extent in comparison to those solutions received in ordinary Einstein-Maxwell or Einstein-Yang-Mills theories.
For instance, their thermodynamical properties are quite unconventional, i.e., extremal dilaton black holes
have zero entropy but its temperature is non-zero \cite{pre91}. Moreover, their properties connected with
late-time behavior of scalar fields in their backgrounds \cite{mod01a,mod01b}, decay of hair 
on them \cite{rog07,rog08}, thermodynamical properties and inequalities among mass, charge and angular
momentum \cite{rog94,rog95,rog00,rog02,rog02b,rog97,rog98,rog14}, uniqueness theorems \cite{rog99,rog,rog10} and dynamical 
collapse process and formation of singularities \cite{bor11,nak12prd,nak12cqg,nak15}, are quite not trivial.
 
In order to have the full insight in the properties of the aforementioned black object it is interesting to
elaborate the problem of a static and dynamical behavior of the simple brane in the spacetime of higher
dimensional dilaton black hole. The first studies of such a problem were presented in \cite{mod03}, where
the existence of a brane (domain wall) was mimicked by the real scalar field with the $\phi^4$ and sine-Gordon
potentials. The previous investigations were restricted to the four-dimensional case and thick branes.
It was revealed that in the case of the extremal dilaton black hole one always obtained an expulsion of the
domain wall from the black hole in question.

Our studies were conducted for three values of the coupling constant $\alpha$, which referred to the uncoupled dilaton field, the low-energy limit of the string theory and dimensionally reduced
Klein-Kaluza theory. It turns out that the value of the dilaton coupling does not influence the static configurations.  However, the brane location in the black object spacetime is connected 
with the dimensionality of the spacetime in question. The brane bend in the nearby of the black object event horizon disappears closer to it as the dimensionality of the bulk increases.
The possible explanation of this fact is connected with the gravity feature which can penetrate the additional dimensions. The same force (bounded with the black hole mass) ought to act
in the additional dimensions and therefore is weaker in comparison to the four-dimensional case. Thus, one can draw a conclusion that, the larger dimensionality of the spacetime manifold one considers, the weaker
force is exerted on the brane in question.
It seems to  be the common feature of the higher-dimensional systems. The very similar behavior
was noticed in studies of the absorption cross section, for massive Dirac field in higher dimensional
black hole spacetime. Namely, $|A|^2$ for the bulk massive Dirac fermions in the 
spacetime of $(n+4)$-dimensional Schwarzschild black hole decreases with the increase of the dimensionality
of the spacetime \cite{rog09}. For the lower energy spectrum we have the same conclusion concerning
the higher dimensional black hole luminosity.


In the case of the extremal dilaton black hole-brane configurations, one always gets expulsion 
of the brane from the black objects,
disregarding dimension of the spacetime and the coupling constant of the theory under consideration. 
This fact confirms the previous results obtained for the four-dimensional dilaton black hole-brane system \cite{mod03}.
In four dimensions the same effect was also revealed for the Abelian Higgs cosmic strings and extremal 
dilaton black holes \cite{mod98,mod99}.

The dynamical evolution of the considered system was elaborated as the series of snapshots
of the temporal evolution of the brane-black hole, for nonextremal and extremal black objects.
The energy density of the scalar field was examined in the black hole spacetime in cylindrical coordinates.
We have considered various spacetime dimensions, i.e., 4,~5,~6,~7 and 8, as well as, the changes of the brane
thickness. In the nonextremal case, it turned out that the black hole is expelled earlier from the
brane in the bulk of a larger dimension and when the brane is thicker. Namely, the radius of the 
area of the brane deformation has a bigger value for a thicker brane. It moves from the initial
position to a larger distance outside the black hole in question.
As far as the energy is concerned, the energy density of the thick brane is smaller than for a thin one.
Moreover, it happens that the bigger initial velocity we examine, the earlier expulsion of the black hole
we observe. The large value of velocity causes the increase of the area of the brane deformation in the
vicinity of the black object.
Similarly to the static
configurations, the dilatonic coupling value does not influence the process in question.

On the other hand, for the extremal dilaton black hole all the aforementioned conclusions are also valid.
The only difference is that the expulsion takes place a little bit earlier than in the nonextremal case.

The current studies could be broadened in several directions, the examples of which are the following. One of them is disregarding the restriction of the codimension-one brane and investigating other values of the bulk spacetime and the brane relative dimensionalities. An assumption of spherical symmetry could be also weakened so that static and stationary axisymmetric spacetimes could be considered.

\acknowledgments

A.N., R.M. and M.R. were partially supported by~the~Polish National Science Centre grant no.~DEC-2014/15/B/ST2/00089. 
{\L}.N. was supported by the Polish National Science Centre under postdoctoral scholarship
FUGA DEC-2014/12/S/ST2/00332.







\end{document}